\documentclass[UTF8,a4paper]{article}
\usepackage{amsmath}
\usepackage{graphicx}
\usepackage{subfigure}
\usepackage{epstopdf}
\usepackage{geometry}
\usepackage{ulem}
\usepackage{indentfirst}
\usepackage{amsfonts,amssymb,dsfont}
\usepackage{setspace}
\usepackage{threeparttable}
\usepackage{amsmath,mathtools,amsthm}
\usepackage{array}
\usepackage{extarrows}
\usepackage{upgreek}

\makeatletter
\renewcommand*\env@matrix[1][\arraystretch]{%
  \edef\arraystretch{#1}%
  \hskip -\arraycolsep
  \let\@ifnextchar\new@ifnextchar
  \array{*\c@MaxMatrixCols c}}
\makeatother
\begin{document}

%QED 3+1: Appelquist T W, Bowick M, Karabali D and Wijewardhana L C R 1986 Phys. Rev. D 33 3704
%Optical evidence for a Weyl semimetal state in pyrochlore Eu 2 Ir 2 O 7

\title{\bf Electronic transport and the related anomalous effects in silicene-like hexagonal lattice}
\author{Chen-Huan Wu
\thanks{chenhuanwu1@gmail.com}
\\Key Laboratory of Atomic $\&$ Molecular Physics and Functional Materials of Gansu Province,
\\College of Physics and Electronic Engineering, Northwest Normal University, Lanzhou 730070, China}

\maketitle
\vspace{-30pt}
\begin{abstract}
\begin{large}
We investigate the anomalous effects due to the Berry correction and the considerable perturbations
in the silicene-like hexagonal lattice system.
The Berry curvature in periodic Bloch band system which related to the electromagnetic field is explored, 
the induced transverse anomalous velocity gives rise to the intrinsic Hall conductivity (without the vertex correction)
expecially in the quantum anomalous Hall phase.
The quantum anomalous Hall effect which related to the anomalous velocity term is detected, including the band avoided corssing effect 
and the generated special band gap.
%{Enlarged Galilean symmetry of anyons and the Hall effect}
The topological spin transport is affected by the Berry curvature
%{Topological spin transport of a relativistic electron}
%{Anomalous direction for skyrmion bubble motion}
and the spin-current-induced Skyrmion spin texture motion is contrasted between the quantum spin Hall effect and quantum anomalous Hall effect.
Since silicene involving the orbital degree of freedom,
the orbital magnetic moment and orbital magnetization contributes significantly to the electronic transport properties of silicene
%{Photoinduced quantum spin and valley Hall effects, and orbital magnetization in monolayer MoS2}
as explored in this article.
We also investigate the electronic tunneling properties of silicene in Josephon junction
with the electric-field-induced Rashba-coupling,
the anomalous effect due to the Berry phase is mentioned.
Our results is meaningful to the application of the spintronics and valleytronics base on the silicene-like topological insulators.

\end{large}

\end{abstract}
\begin{large}
\section{Introduction}

%electrostatic potential=gate voltage
In this paper, we investigate the anomalous effects induced by the Berry correction (semiclassical correction) to the electronic transport (tunneling)
of the non-relativistic electrons (or quasiparticles) in silicene.
The non-relativistic case origin from the small Fermi velocity compared to the speed of light,
which is $v_{F}=5.5\times 10^{5}\approx c/500$ for silicene and smaller than that of graphene which is $v_{F}\approx c/300$.
There is also a distinction between the relativistic and non-relativistic case:
The main parameter during the semiclassical motion of particles is momentum (quasimomentum) for non-relativistic case,
while it is wavelength for relativistic case.
%{Topological spin transport of a relativistic electron}
Due to the semiclassical correction to the equation of motion of particles,
the trajectory of particle is distincted from the traditional one,
especially when it's under an electromagnetic field or a circular light field,
which are related to the Berry curvature (Berry gauge field).
Moreover, artificial gauge field has been successfully created base on the non-Abelian Berry phase in the nonadiabatic case\cite{Zhu S L}.
The unconventional semiclassical motion with nonzero Berry curvature also leads to the topological nontrivial
spin transport as well as the intriguing spin/valley Hall conductivity in the topological insulators like silicene or MoS$_{2}$ or germanene.
The topological spin transport as well as the momentum of center-of-mass of the wave package 
are affected by the applied electric field, magnetic field, and the off-resonance circularly polarized light.
The anomalous velocity due to the Berry curvature
(including a Lorentz-like term) shifts the electrons in the direction transverse to the electric field and magnetic field
(which is also the direction of the spin accumulation by the spin Hall effect\cite{Ummelen F C}),
%{Coupled Spin and Valley Physics in Monolayers of MoS2 and Other Group-VI Dichalcogenides}
and gives rise to the spin or valley (transverse) Hall conductivity,
while the circularly polarized light results in the opposite-spin configurartion of the neighbor valleys\cite{Xiao D2}
which will affects the spin polarization and charge current,
%{Anomalous direction for skyrmion bubble motion}
and the motion of electrons is along the applied fields which can be obtained through the expression of $\hbar\partial_{t}{\bf k}$
as we state in this article.
The electron and hole are be endowed with the velocities in opposite direction and thus give rise to the Hall conductivity in the presence of 
longitudinal charge current.
The spin accumulation caused by spin current in quantum spin Hall phase is orthogonal to the electric field and the charge current,
note that the spin current here is unlike the one caused by the conduction electron flow which is along the direction of applied electric field,
thus it's recently observed\cite{Ummelen F C} experimentally for the anomalous motion of skyrmion carrired by the spin current in quntum spin Hall phase.
Base on the optical spin-valley-coupled selection rule, this spin accumulation can be facilitated by the applied circularly polarized light
(the so-called photo-current-induced spin accumulation\cite{Endres B}) or the magneto-optic Kerr effect.
%{Coupled Spin and Valley Physics in Monolayers of MoS2 and Other Group-VI Dichalcogenides}

The largest difference between the effective Dirac Hamiltonian in relativistic case and non-relativistic case is the emergement of the Zeeman-like exchange field
term and the (intrinsic and external) spin-orbit coupling (SOC) term,
and the classical mass is replaced by the Dirac-mass term (or the mass about the interlayer hopping and intralayer hopping for the bilayer silicene\cite{Wu C H5}).
It has also been proved experimentally\cite{Morpurgo A F} early that the SOC-induced Berry 
geometric phase affects deeply the quantum transport (like the spin or valley) just like the Aharonov-Anandan geometric phase induced by magnetic field.
%{Nonadiabatic Noncyclic Geometric Phase and Ensemble Average Spectrum of Conductance in Disordered Mesoscopic Rings with Spin-Orbit Coupling.pdf}
The intrinsic spin or valley Hall conductivity (or polarization)\cite{Tahir M} is also related to the Berry curvature which induce the anomalous electron motion
under the electromagnetic field. Intriguingly, the intrinsic quantum spin Hall phase in silicene can be changed to the trivial insulator 
with obvious variance of conductivity under high external gate voltage\cite{Dyrda A}.
The anomalous velocity term also related to the quantum anomalous Hall effect\cite{Horváthy P A},
which with a special nonzero Chern number and an anticrossing band-induced gap as we explored in this paper.
We also found that the anomalous Hall effect can be generated by the exchange field and the electric field.
The Berry phase is just like the Aharonov-Bohm phase induced by the magnetic field\cite{Chen J W,Zhu S L},
while for the case that only have the electric field,
it cause a topological spin transport which induce a spin current\cite{Bliokh K Y} like the quantum spin Hall effect.
%{Topological spin transport of a relativistic electron}

Distinct to the Dirac fermions (nonrelativitic when $m_{D}^{\eta\sigma_{z}\tau_{z}}\neq 0$) in the low-energy tightbing midel as shown in Appendix.A,
the Bulk effective Hamiltonian of the weyl semimetal or three-dimension topological insulator requires the nonzero longitudinal momentum $k_{z}$
and quadradic dispersion $k^{2}$ which for silicene or graphene (hexagonal quantum spin Hall materials) can emerges only for bilayer or multilayer form.
Here the linear $k_{z}$-dependence rised with the increase of overlap between the $p_{z}$ orbit of atoms,
%{Signatures of bulk topology in the non-linear optical spectra of Dirac-Weyl materials}
%{Bulk and interface quantum states of electrons in multi-layer heterostructures with topological materials}
%{Bloch-Siegert Shift in Dirac-Weyl Fermionic Systems }
in the case of $k_{x}=k_{y}=0$.
In the mean time, nonzero Berry curvature (like the case of quantum anomalous Hall effect) also leads to the anomalous 
velocity\cite{Duval C,Stone M} (in semiclassical correction of solid),
and it's related to the Anomalous Rabi oscillation in the Dirac-Weyl Fermionic systems\cite{Kumar U}
as well as the dispersion of the Andreev bound state\cite{Wu C H7,Wu C H5,Deacon R S}.
The anomalous Rabi oscillation frequency is unlike the conventional Rabi frequency, it’s
related to the Chern number: When the Chern number of silicene (or for other topological
insulator system) is zero, then the induced (by anomalous Rabi oscillation frequency) mass is
nonzero for this trivial system and thus with the gapped edge states; while when the Chern
number is nonzero, the induced mass is zero and corresponds to the non-trivial system which
with the gapless edge states.
While for the Andreev bound states,
It becomes gapless for the topological state
(or with an infinitesimal gap like the topological band with nonzero Chern number).
%{Transport properties of nonequilibrium systems under the application of light Photoinduced quantum Hall insulators without Landau levels}
Consider the amplitude of the envelope function of the two sublattices,
the quasienergy of silicene can also be described by the Weyl function as
$\varepsilon=\gamma({\pmb \sigma}\cdot{\bf k})$,
%{berry's phase and absence of back scattering in carbon nanotubes}
%{Band Structure of Graphite  J. C. SLoNczEwsKIt AND P. R. Wzrss 1957}
with the band parameter $\gamma=2{\bf k}_{F}/(\pi\Pi({\bf q},0)\hbar)$($\sim v_{F}$ for monolayer silicene with large carriers density,
and $\Pi({\bf q},0)$ is the dynamical polarization) 
%{Dielectric function, screening, and plasmons in two-dimensional graphene}
which is inversely proportional to the static polarization function.

\section{Low-energy tight-binding model}

The time-dependent Dirac Hamiltonian in tight-binding model of the monolayer silicene under time-dependent vector potential reads\cite{Wu C H2,Wu C H7}
%Ref.\cite{Ezawa M} as
\begin{equation} 
\begin{aligned}
H(t)=&\hbar v_{F}(\eta\tau_{x}P_{x}(t)+\tau_{y}P_{y}(t))+\eta\lambda_{{\rm SOC}}\tau_{z}\sigma_{z}+a\lambda_{R_{2}}\eta\tau_{z}(P_{y}\sigma_{x}-P_{x}\sigma_{y})\\
&-\frac{\overline{\Delta}}{2}E_{\perp}\tau_{z}+\frac{\lambda_{R_{1}}}{2}(\eta\sigma_{y}\tau_{x}-\sigma_{x}\tau_{y})+M_{s}s_{z}+M_{c}
-\eta\tau_{z}\hbar v_{F}^{2}\frac{\mathcal{A}}{\Omega}+\mu,
\end{aligned}
\end{equation}
where 
$P_{x}(t)=k_{x}-eA_{x}(t)=k_{x}-eA{\rm sin}\Omega t$.
$E_{\perp}$ is the perpendicularly applied electric field, 
$a=3.86$ is the lattice constant,
$\mu$ is the chemical potential,
$\overline{\Delta}$ is the buckled distance between the upper sublattice and lower sublattice,
$\sigma_{z}$ and $\tau_{z}$ are the spin and sublattice (pseudospin) degrees of freedom, respectively.
$\eta=\pm 1$ for K and K' valley, respectively.
$M_{s}$ is the spin-dependent exchange field and $M_{c}$ is the charge-dependent exchange field.
$\lambda_{SOC}=3.9$ meV is the strength of intrinsic spin-orbit coupling (SOC) and $\lambda_{R_{2}}=0.7$ meV is the intrinsic Rashba coupling
which is a next-nearest-neightbor (NNN) hopping term and breaks the lattice inversion symmetry.
$\lambda_{R_{1}}$ is the electric field-induced nearest-neighbor (NN) Rashba coupling which has been found that linear with the applied electric field
in our previous works\cite{Wu C H1,Wu C H2,Wu C H3,Wu C H4,Wu C H5}, which as $\lambda_{R_{1}}=0.012E_{\perp}$.
Note that we ignore the effects of the high-energy bands on the low-energy bands.
%{Photoinduced quantum spin and valley Hall effects, and orbital magnetization in monolayer MoS2}

Due to the perpendicular electric field $E_{\perp}$ and the off-resonance circularly polarized light which with frequency $\Omega\gg 1000$ THz,
the Dirac-mass and the corresponding quasienergy spectrum (obtained throught the diagonalization procedure)
%{Photoinduced quantum spin and valley Hall effects, and orbital magnetization in monolayer MoS2}
%{Photoinduced pseudospin effects in silicene beyond the off-resonant condition}
%{Spin valleytronics in silicene_ Quantum spin Hall–quantum anomalous Hall insulators and single-valley semimetals}
%{Valley-Polarized Metals and Quantum Anomalous Hall Effect in Silicene}
%{Photoinduced Topological Phase Transition and a Single Dirac-Cone State in Silicene}
%{Superconducting proximity effect in silicene- Spin-valley-polarized Andreev reflection, nonlocal transport, and supercurrent}
\begin{equation} 
\begin{aligned}
&m_{D}^{\eta\sigma_{z}\tau_{z}}=|\eta\sqrt{\lambda_{{\rm SOC}}^{2}+a^{2}\lambda^{2}_{R_{2}}k^{2}}s_{z}\tau_{z}-\frac{\overline{\Delta}}{2}E_{\perp}\tau_{z}+M_{s}s_{z}-\eta\hbar v_{F}^{2}\frac{\mathcal{A}}{\Omega}|,\\
&\varepsilon=s\sqrt{a^{2}\lambda^{2}_{R_{2}}k^{2}+(\sqrt{\hbar^{2}v_{F}^{2}{\bf k}^{2}
+(\eta\lambda_{{\rm SOC}}s_{z}\tau_{z}-\frac{\overline{\Delta}}{2}E_{\perp}\tau_{z}-\eta\hbar v_{F}^{2}\frac{\mathcal{A}}{\Omega} )^{2}}+M_{s}s_{z}+s\mu)^{2}},
\end{aligned}
\end{equation}
respectively, where the dimensionless intensity $\mathcal{A}=eAa/\hbar$ is in a form similar to the Bloch frequency, 
and $s=\pm 1$ is the electron/hole index, and the
subscript $e$ and $h$ denotes the electron and hole, respectively.
Note that here this Dirac-mass is correct for exchange field $|M_{s}|\le\lambda_{SOC}(1+\frac{a^{2}\lambda^{2}_{R_{2}}}{\hbar^{2}v_{F}^{2}})$,
and the resonance will be presented in the following.
%{Valley-Polarized Metals and Quantum Anomalous Hall Effect in Silicene}
The off-resonance circularly polarized light results in the asymmetry band gap in two valleys (see Ref.\cite{Wu C H5})
and breaks the time-reversal symmetry in the mean time,
and thus provides two pairs of the different incident electrons that may leads to the josephson current reversal
due to the valley-polarization.
The effects of the exchange-field-term $M_{s}s_{z}$ and $M_{c}$ are presented in the Fig.1(a)-(c),
where we can see that the $M_{s}$ which is spin-dependent generally close the gap
and shift the band with up-spin and down-spin upward and downward, respectively, while the charge-dependent $M_{c}$ just move the whole band tructure upward
but does not breaks the spin degenerate which shows it may related to the valley degree of freedom $\eta$.
It's obviouly that the avoided corssing effect emerges and it
is enhanced with the increase of electric field (or magnetic field),
that also contributes to the Chern number by the Skyrmion
spin texture as we discuss below. 
Here we note that the motion of the Skyrmion spin texture here carried by the spin current due to the angular momentum conservation\cite{Ummelen F C} is rather weak
than the one in the bulk sample,
and the Skyrmion is also much more stable in the quantum anomalous Hall phase which with chiral edge 
than that in the quantum spin Hall phase
which with helical edge.
%{Anomalous direction for skyrmion bubble motion}
The avoided corssing effect also observable in the Floquet system under the electromagnetic field\cite{Perez-Piskunow P M}.
%{Hierarchy of Floquet gaps and edge states for driven honeycomb lattices}
%{Large tunability of lattice thermal conductivity of monolayer silicene via mechanical strain}
For the band structure that both the intrinsic SOC and the NNN electric field-induced Rashba-coupling $\lambda_{R_{1}}$ are taken into account,
the symmetry between the conduction band and valence band is broken, as shown in Fig.1(d),
and the band splitting between the two conduction bands is decreased while that of the two valence bands is increased.
Due to the spin mixing by the extrinsic Rashba-coupling $\lambda_{R_{1}}$,
the band splitting here is no more the spin-splitting but related to the index $(\alpha,\beta)$,
\begin{equation} 
\begin{aligned}
\varepsilon=s\sqrt{a^{2}\lambda^{2}_{R_{2}}k^{2}+(\alpha\lambda_{R_{1}}+\beta\sqrt{\hbar^{2}v_{F}^{2}{\bf k}^{2}
+(\eta(\lambda_{R_{1}}-\alpha\lambda_{{\rm SOC}})s_{z}\tau_{z}-\frac{\overline{\Delta}}{2}E_{\perp}\tau_{z}-\eta\hbar v_{F}^{2}\frac{\mathcal{A}}{\Omega} )^{2}}+M_{s}s_{z}+s\mu)^{2}},
\end{aligned}
\end{equation}
%{Dielectric function, screening, and plasmons of graphene in the presence of spin-orbit interactions}
%{Giant spin rotation under quasiparticle-photoelectron conversion Joint effect of sublattice interference and spin-orbit coupling}
%{Reversing Berry phase and modulating Andreev reflection by Rashba spin-orbit coupling in graphene mono- and bilayers.pdf}
%{Tunable anomalous Andreev reflection and triplet pairings in spin-orbit-coupled graphene}
Through Fig.1(d), we can see that the energy splitting of band structue is controlled by the index $(\alpha,\beta)$,
and the configuration of spin helical is symmetry between the conduction band and valence band.
Such phenomenon also emerges for the case of off-resonance light to the MoS$_{2}$\cite{Tahir M,Xiao D2,Scholz A} 
which has a much larger intrinsic SOC than graphene or silicene,
and in the mean time, due to the $\eta$-dependent optical term in Eq.(1),
the valley asymmetry is rised (see \cite{Wu C H2,Wu C H5}) and results in the possible 100$\%$ valley polarization with the almost pure
valley transport like a valley filter.
%{Valley filter and valley valve in graphene}
The reduction of the Dirac-mass is also accompanied by the rise of the longitudinal conductivity\cite{Tahir M}.
In addition, the off-resonance light also enhance the difference of the orbital magnetic moment between two inequivalent valleys,
as we discuss in following text.

In the presence of vertical electric field as well as the first-order and second-order Rashba-coupling,
the system of silicene can be described by Hamiltonian
$H=\Psi^{\dag}H^{\pm}_{{\rm eff}}\Psi/2$ in the low-energy Dirac theory, with 
the two-component 
spinor-valued field operators 
\begin{equation} 
\begin{aligned}
\Psi=[(\psi_{\uparrow}^{A},\psi_{\downarrow}^{A},\psi_{\uparrow}^{B},\psi_{\downarrow}^{B}),
((\psi_{\uparrow}^{A\dag},\psi_{\downarrow}^{A\dag},\psi_{\uparrow}^{B\dag},\psi_{\downarrow}^{B\dag}))]^{T}.
\end{aligned}
\end{equation}
The BCS-like effective Hamiltonians $H_{{\rm eff}}$ in the basis of $\{\tau\otimes\sigma\}$ read
\begin{equation} 
\begin{aligned}
H_{{\rm eff}}^{K}=
\begin{pmatrix}
 m_{D}^{+ ++}&\hbar v_{F}(k_{x}-ik_{y})&ia\lambda_{R_{2}}(k_{x}-ik_{y})&0\\
 \hbar v_{F}(k_{x}+ik_{y})    &m_{D}^{+ +-} &-i\lambda_{R_{1}}&-ia\lambda_{R_{2}}(k_{x}-ik_{y})\\
-ia\lambda_{R_{2}}(k_{x}+ik_{y})   &i\lambda_{R_{1}} &m_{D}^{+ -+}&\hbar v_{F}(k_{x}-ik_{y})\\    
0& ia\lambda_{R_{2}}(k_{x}+ik_{y}) &\hbar v_{F}(k_{x}+ik_{y})&m_{D}^{+ --}\\  
\end{pmatrix},
\end{aligned}
\end{equation}
\begin{equation} 
\begin{aligned}
H_{{\rm eff}}^{K'}=
\begin{pmatrix}
 m_{D}^{- ++}&\hbar v_{F}(k_{x}+ik_{y})&ia\lambda_{R_{2}}(k_{x}-ik_{y})&-i\lambda_{R_{1}}\\
 \hbar v_{F}(k_{x}-ik_{y})    &m_{D}^{- +-} &0&ia\lambda_{R_{2}}(k_{x}-ik_{y})\\
ia\lambda_{R_{2}}(k_{x}+ik_{y})   &0 &m_{D}^{- -+}&\hbar v_{F}(k_{x}+ik_{y})\\    
-i\lambda_{R_{1}}& -ia\lambda_{R_{2}}(k_{x}-ik_{y}) &\hbar v_{F}(k_{x}+ik_{y})&m_{D}^{- --}\\  
\end{pmatrix}.
\end{aligned}
\end{equation}
While for the bilayer silicene with the NN interlayer hopping $t'=2$ eV
%{which we consider the AB-stacked bilayer silicene here}
and we ignore the NNN interlayer hopping.
The interlayer SOC is estimated as 0.5 meV here\cite{Ezawa M2} and since the 
trigonal warping term between two layers has a non-negligible impact when apply the off-resonance light in terahert range\cite{Morell E S},
we set the trigonal warping hopping parameter as $t_{w}=0.16$ eV here.
Then the low-energy Dirac effective model can be written in a matrix form\cite{Wu C H5}:
\begin{equation} 
\begin{aligned}
H_{{\rm eff}}^{bi}=\eta\begin{pmatrix}m_{D}^{\eta ++}-\eta\hbar v_{F}^{2}\frac{\mathcal{A}^{2}\Omega}{t^{'2}}&\hbar v_{w}(k_{x}+ik_{y})&0&\hbar v_{F}(k_{x}-ik_{y})\\
 \hbar v_{w}(k_{x}-ik_{y})    &m_{D}^{\eta +-}+\eta\hbar v_{F}^{2}\frac{\mathcal{A}^{2}\Omega}{t^{'2}}&\hbar v_{F}(k_{x}+ik_{y})&0\\
 0&\hbar v_{F}(k_{x}-ik_{y})   &m_{D}^{\eta -+}&\eta t'\\        
 \hbar v_{F}(k_{x}+ik_{y})   &0&\eta t'&m_{D}^{\eta --}\\  
 \end{pmatrix},
\end{aligned}
\end{equation}
where $v_{w}=\sqrt{3}at_{w}/2\hbar$ is the velocity associates with the trigonal warping\cite{Wu C H5}.
The valley symmetry is broken due to the trigonal warping term here,
which may leads to the single-Dirac-cone state,
that implies that the light in a finite intensity has the same effect with the out-of-plane antiferromagnetic exchange field.
%The asymmetry effect on the single valley band structure can be seen in Fig.1(b),
%where we consider the NN Rashba coupling but ignore the exchange field.
%{Quasi-Topological Insulator and Trigonal Warping in Gated Bilayer Silicene}
%{Radiation effects on the electronic properties of bilayer graphene}
Since the time-reversal-invariance is broken by the off-resonance light, the 
valley symmetry is broken and the up-spin and down-spin flow with different velocities in each edge,
and that is the fundament of the realization of single-Dirac cone state\cite{Wu C H5}.
%{Photoinduced Topological Phase Transition and a Single Dirac-Cone State in Silicene}

Through the procedure of block diagonalization as mentioned in Ref.\cite{López A},
%{Quantum anomalous Hall effect in a flat band ferromagnet}
the above matrices can be simplified as\cite{Wu C H5}
\begin{equation}
\begin{aligned}
H_{{\rm eff}}=
\begin{pmatrix}
m_{D}^{\eta s_{z}\tau_{z}} &\hbar v_{F}(k_{x}-ik_{y})e^{-i\theta}\\ 
\hbar v_{F}(k_{x}+ik_{y})e^{i\theta}& -m_{D}^{\eta s_{z}\tau_{z}}
\end{pmatrix},\\
{\rm with}\\
\hbar v_{F}{\bf k}=\hbar v_{F}\begin{pmatrix}0 &k_{x}-ik_{y}\\ k_{x}+ik_{y}& 0\end{pmatrix}
=\pm t|1+e^{ik_{x}}+e^{-ik_{y}}|,
%{\rm and},
%m_{D}^{*\eta\sigma_{z}\tau_{z}}=|\eta\sqrt{\lambda_{{\rm SOC}}^{2}+a^{2}\lambda_{R_{2}}^{2}}s_{z}\tau_{z}
%-\frac{\overline{\Delta}}{2}E_{\perp}\tau_{z}+M_{s}s_{z}+M_{c}+\lambda_{R_{1}}/2|.
\end{aligned}
\end{equation}
where $\theta$ is the angle between ${\bf k}$ and $k_{x}$.
$k_{x}+ik_{y}=\vec{\sigma}_{AB}\cdot{\bf k}$\cite{Kumar U2,Kumar V}.
%with the vector $\vec{\sigma}_{AB}=\sigma_{z}\cdot({\rm cos}(E_{ex}\Phi),{\rm sin}(E_{ex}\Phi),0)$
%describes the adiabatic evolution with the variable $\Phi$ and the induced frustrations.
%For pseudospin, it's usually defined as $\Phi=\phi={\rm arctan}\frac{k_{y}}{k_{x}}$.
%$E_{ex}$ is the exchange field which related to the chirality.
Here we comment that the exchange field for the out-of-plane polarization is much larger than the in-plane one.
Note that we don't consider the $\lambda_{R_{1}}$ and $\lambda_{R_{2}}$ term in the above Dirac-mass.
%if they are contained, the Dirac-mass becomes
%\begin{equation} 
%\begin{aligned}
%m_{D}^{'\eta\sigma_{z}}=|\eta\sqrt{\lambda_{{\rm SOC}}^{2}+a^{2}\lambda_{R_{2}}^{2}}s_{z}\tau_{z}
%-\frac{\overline{\Delta}}{2}E_{\perp}\tau_{z}+Ms_{z}-\eta\hbar v_{F}^{2}\frac{\mathcal{A}}{\Omega}+\lambda_{R_{1}}/2|.
%\end{aligned}
%\end{equation}
While for the bilayer silicene, the simplified matrix becomes
\begin{equation}
\begin{aligned}
H_{{\rm eff}}^{bi}=
\begin{pmatrix}
m_{D}^{\eta s_{z}\tau_{z}} &\frac{\hbar^{2}}{2m^{*}} (k_{x}-ik_{y})^{2}e^{-2i\theta}\\ 
\frac{\hbar^{2}}{2m^{*}} (k_{x}+ik_{y})^{2}e^{2i\theta}& -m_{D}^{\eta s_{z}\tau_{z}}
\end{pmatrix},\\
{\rm with}\\
\frac{\hbar^{2}{\bf k}^{2}}{2m^{*}}=\frac{{\bf k}^{2}t^{2}}{t'}
=\frac{\hbar^{2}}{2m^{*}}\begin{pmatrix}0 &(k_{x}-ik_{y})^{2}\\ (k_{x}+ik_{y})^{2}& 0\end{pmatrix}=
\pm\frac{t'}{2}\pm\sqrt{(\frac{t'}{2})^{2}+t^{2} |1+e^{ik_{x}}+e^{-ik_{y}}|^{2}},
\end{aligned}
\end{equation}
%{Berry phase and pseudospin winding number in bilayer graphene}
%{Unconventional quantum Hall effect and Berry's phase of 2 in bilayer graphene}
where $t'$ is the interlayer hopping, and
that can be easily deduced by $m^{*}=\hbar^{2}t'/(2t^{2})\sim 1/v_{F}^{2}$ and $\hbar v_{F}=\frac{\sqrt{3}}{2}at$.
Note that the $m^{*}$ here is much smaller than the free electron mass, 
and it's related to the interlayer and intralayer hopping and
the velocity of these hopping is much slower than the speed of light thus give rises the non-relativity effect
as we introduced at the begining.
Due to the possible quadratic dispersion in the energy band bottom,
the momentum-independent effective mass reads $m^{*}_{bottom}=\frac{4m_{D}^{\eta \sigma_{z}\tau_{z}}}{\hbar^{2}v_{F}^{2}}$.
%{Valley-Contrasting Physics in Graphene Magnetic Moment and Topological Transport}
The above expression results in the four band structure in the spin degenerate case for the silicene bilayer.

The momentum above can be replaced by the canonical (covariant) momentum through the minimal substitution as $k_{x}\rightarrow P_{x}-\frac{e}{c}A_{x}$
where $A_{x}$ is the $x$-component of the vector potential ${\bf A}$.
For relativitic particle, the canonical momentum satisfies $P=-i\hbar\partial_{r}c$ since $\partial_{r}=-\frac{1}{c}\partial_{t}$\cite{Bliokh K Y},
and it's useful in the block diagonalization of Dirac equation as well as the use of BMT equation.
%{Semiclassical limit of the Dirac equation and spin precession}
Since we applying the magnetic field perpendicular to the silicene, i.e.,
${\bf B}=\nabla\times{\bf A}=(0,0,B_{z})$,
with the Landau gauge ${\bf A}=(-B_{z}y,0,0)$,
%{Quantum Hall effects in a Weyl semimetal Possible application in pyrochlore iridates}
%{Integer and half-integer quantum Hall effect in silicene Influence of an external electric field and impurities}
%{Anomalous orbital magnetism in Dirac-electron systems Role of pseudospin paramagnetism}
%{Anomalous integer quantum Hall effect in A A-stacked bilayer graphene}
%{Topological spin transport of a relativistic electron}
and since the momentum can be replaced by the covariant one in Peierls phase,
the ladder operators satisfy $[{\bf P},{\bf P}^{\dag}]=1$, 
%{Chiral anomaly enhancement and photoirradiation effects in multiband touching fermion systems}
with ${\bf P}=\hbar(\frac{y}{\ell_{B}}-\ell_{B}k_{x}+\partial_{y})$ and ${\bf P}^{\dag}=\hbar(\frac{y}{\ell_{B}}-\ell_{B}k_{x}-\partial_{y})$.

\section{Berry curvature with external electromagnetic field}

In the presence of scattering by the charged impurity (spin-orbit scattering)
with a scattering potential larger than the lattice constant\cite{Ando T}
and the strong SOC (compared to the graphene or black phosphorus),
%{berry's phase and absence of back scattering in carbon nanotubes}
%{Magnetoresistance in two-dimensional disordered systems: effects of Zeeman splitting and spin-orbit scattering}
the elastic back-scattering which with the conserving spin is suppressed
since the spin rotates as the wave vector changes direction during the scattering (in fact, the spin is always in the direction of the wave vector
due to the helicity of the dirac equation in the case of time-reversal invariant).
In the presence of time-reversal-invariance, the spin-momentum locking can be observed.
%{Spontaneous supercurrent and ?0 phase shift parallel to magnetized topological insulator interfaces}
That also implies that, for the quadratic edge state dispersion,
the back-scattering is only possible in low-energy region where the difference between spin directions is $\ll 180^\text{o}$.
%{Testing Topological Protection of Edge States in Hexagonal Quantum Spin Hall Candidate Materials}
As the Bloch wave vector undergoes a rotation around the whole pseudospin space,
%{berry's phase and absence of back scattering in carbon nanotubes}
%{Berry phase and pseudospin winding number in bilayer graphene}
%{Unconventional quantum Hall effect and Berry's phase of 2π in bilayer graphene}
the Berry phase of monolayer silicene which is gauge invariant 
(in contrast to the anomalous case with the perturbations) can be obtained by the integral of the time within the period of a full rotation
(a loop winding around the K-point)
%{Unconventional states of bosons with the synthetic spin–orbit coupling}
\begin{equation}
\begin{aligned}
\Gamma=-i\lim_{\phi\rightarrow 0}\int^{\frac{1}{2}(1+e^{i\phi})}_{1}\frac{1}{t}dt \frac{2\pi}{\phi}=\pi,
\end{aligned}
\end{equation}
with the position-dependent phase $\phi={\rm arctan}\frac{{\bf k}_{i+1}-{\bf k}_{i}}{{\bf k}_{i}}$.
For bilayer silicene, the Berry phase can be obtained as $2\pi$ through the simialr procedure.
In fact, for the adiabatic case without the perturbations, the trivial Berry phase is constant and the $e^{i\Gamma}$ is gauge-invariant,
%{Berry phase and pseudospin winding number in bilayer graphene}
which is equivalent to the symmetry case.
Note that here we don't consider the SOC, and thus the gauge invariance is obtained through the winding number of pseudospin in momentum space.
%{Unconventional quantum Hall effect and Berry's phase of 2π in bilayer graphene}
%{Berry phase and pseudospin winding number in bilayer graphene}
%{Reversing Berry phase and modulating Andreev reflection by Rashba spin-orbit coupling in graphene mono- and bilayers}
While the SOC will slightly reduce the Berry phase in monolayer silicene and enlarge the Berry phase in bilayer silicene.

For the wave vector rotates as a function of time in an adiabatic way,
${\bf k}(t=0)={\bf k}(t=T)$,
the berry curvature, which well describe the local property of the band
%{Mapping the Berry curvature from semiclassical dynamics in optical lattices}
structure for adiabatic transport, can be obtained as a triple integral with the second-rank tensor field, 
$\Omega_{i}({\bf k})=-{\rm Im}[\varepsilon_{i\mu\nu}\partial_{k\mu}\langle\psi|\partial_{k\nu}\rangle]$
where $\psi$ is the Bloch band state whose period is $L$, $\varepsilon_{i\mu\nu}$ is the Levi-Civit{\`a} tensor.
%period part of bloch function
Using ${\bf I}=\frac{L}{2\pi}\sum_{\psi'}|\psi'\rangle\langle\psi'|$
(here we assume that the 
silicene is subjected to an one-dimension periodic potential along the zigzag direction 
with the period 
$L=2\pi$ thus this normalization term can be omitted), 
%{Anomalous Bloch oscillations in one-dimensional parity-breaking periodic potentials}
$\langle\psi'|\partial_{k\nu}\psi\rangle(\varepsilon_{\psi}-\varepsilon_{\psi'})
=\langle \psi'|\partial_{k\nu}H_{k}|\psi\rangle$, then
the well known Berry curvature formular can be obtained as
%{J. K. Asb′oth, L. Oroszl′any, and A. P′alyi, arXiv:1509.02295}
%{D. Xiao, M-Ch Chang, and Q. Niu, Rev. Mod. Phys. 82, 1959.}
%{http://theorie.physik.uni-konstanz.de/burkard/sites/default/files/ts15/Berry-phase.pdf Berry phase, Chern number}
%{Berry phase correction to electron density in solids and exotic dynamics}
%{ M. V. Berry, Proc. Roy. Soc. Lond. A 392, 45 (1984).}
\begin{equation}
\begin{aligned}
\Omega_{i}({\bf k})=-{\rm Im}\left[ \sum_{\psi'\neq\psi}\frac{\langle\psi'|\partial_{k\mu}H_{k}|\psi\rangle\times\langle\psi'|\partial_{k\nu}H_{k}|\psi\rangle}{(\varepsilon_{\psi}-\varepsilon_{\psi'})^{2}}\right].
\end{aligned}
\end{equation}
Here the Bloch wave function is related to the eigenfunction of the system Hamiltonian by $\psi({\bf r})=e^{-i{\bf k}\cdot{\bf r}}\Psi({\bf r})$.
%{Anomalous Bloch oscillations in one-dimensional parity-breaking periodic potentials}
%{Berry Phase, Hyperorbits, and the Hofstadter Spectrum}
%{Magnetothermoelectric transport properties in graphene superlattices with one-dimensional periodic potentials}
The sum of Berry curvature around the curl is constant thus it's divergence-free and obeys $\partial_{k}\langle\psi|\psi\rangle=0$,
thus $\partial_{k}\Omega({\bf k})=0$ for the Bloch band which does not degenerate with other bands in momentum space.
%{Phase transition between the quantum spin Hall and insulator phases in 3D: emergence of a topological gapless phase}
By using the Berry connection in low energy level, 
%{Berry phase effects on electronic properties}
${\bf A}({\bf k})=i\langle\psi|\partial_{k}\psi\rangle=-{\rm Im}\langle\psi|\partial_{k}\psi\rangle$,
the Berry curvature can be rewritten as
$\Omega_{i}({\bf k})=\varepsilon_{i\mu\nu}\partial_{k\mu}{\bf A}_{\nu}({\bf k}) \hat{s}=\frac{{\bf k}}{k^{3}}\hat{s}
=\frac{1}{k^{2}}\hat{s}\cdot{\bf e_{k}}$ where ${\bf e_{k}}$ labels the radiation direction of the wave vector.
%{Unconventional states of bosons with the synthetic spin–orbit coupling}
%{Chiral anomaly enhancement and photoirradiation effects in multiband touching fermion systems}
%{Berry phase and anomalous velocity of Weyl fermions and Maxwell photons}
where $\hat{s}$ denotes the spin operator as well as its helicity and it's an important indicator 
for the system under magnetic field or light field, and it can also be replaced by a pseudospin operator.
%{Chiral anomaly enhancement and photoirradiation effects in multiband touching fermion systems}
Here we note that, distincted from the case of free pseudospin-1 Maxwell particle, for the massless two-dimension Dirac Fermions near the Dirac cone
where the Berry curvature has obvious peak, the coupling between the magnetic moment and the magnetic field
will leads to a magnetic field-induced energy shift which cancer the effect of the Berry curvature-induced energy-shift\cite{Stone M}
in the case of only the magnetic field exist.
While for the internal magnetic moment for a magnetism matter, it can be directly coupled to the supercurrent for a Josephson device\cite{Buzdin A}.
The
momentum-pseudospin
space Berry curvature in unit of $e^{2}/h$ is in a similar distribution with the orbital magnetic moment
which couples to the magnetic field in $z$-direction
and reads
\cite{Xiao D} 
\begin{equation} 
\begin{aligned}
m({\bf k})=&\frac{e}{\hbar}\varepsilon({\bf k}) \Omega({\bf k})\\
=&\frac{e}{2\hbar}\frac{2\eta \hbar^{2}v_{F}^{2}m^{\eta s_{z}\tau_{z}}_{D}}{2(4(m^{\eta s_{z}\tau_{z}}_{D})^{2}+\hbar^{2}v_{F}^{2}k^{2})}.
\end{aligned}
\end{equation}
%{Valley-contrasting physics in graphene: magnetic moment and topological transport}
%{Electrical tuning of valley magnetic moment through symmetry control in bilayer MoS2}
We can see that the orbital magnetic moment is opposite in sign for two valleys,
which is required by the time-reversal invariant and the accompanied spin-momentum locking.
%{Giant magnetoresistance and perfect spin filter in silicene, germanene, and stanene}
Note that the relation $\hbar v_{F}=\frac{\sqrt{3}}{2}at$ is used here.
The orbital magnetic moment can be probed by X-ray circular dichroism or other electron probes\cite{Boeglin C}.
%{Detecting magnetic ordering with atomic size electron probes}
%{Thickness-dependent magnetic properties and strain-induced orbital magnetic moment in  thin films}
The Berry curvature thus can be expressed as
\begin{equation} 
\begin{aligned}
\Omega({\bf k})
=\frac{1}{2 }\frac{2\eta \hbar^{2}v_{F}^{2}m^{\eta s_{z}\tau_{z}}_{D}}{2\varepsilon(4(m^{\eta s_{z}\tau_{z}}_{D})^{2}+\hbar^{2}v_{F}^{2}k^{2})}.
\end{aligned}
\end{equation}
The orbital magnetic moment is shown in Fig.2 for two different Dirac-mass.
The orbital magnetic moment decays faster for smaller Dirac-mass,
and through the above expression we can obtain that the Berry curvature has the same law. 
It is worth noting that
the nonzero Berry curvature and orbital magnetic moment require the broken symmetry\cite{Wu S}.
Here we list some possible symmetry-broken case in our system:
(I) the Rashba-coupling and perpendicular electric field breaks the inversion symmetry; 
and the spatial inversion symmetry can be broken by the buckled structure like in the GaAs;
%{Enlarged Galilean symmetry of anyons and the Hall effect}
(II)
the off-resonance 
light breaks the time-reversal-invariance (or due to the competition between Zeeman coupling and Rashba-coupling\cite{Meijer F E});
(III) the Rashba-coupling also breaks the chiral symmetry as well as the symmetry between conduction band and valence band
(when in the absence of exchange field);
(IV) the magnetic-field-induced shift (according to L{\"o}wdin perturbation theory) in quasimomenta space breaks the reflection symmetry.
Except that, the distortions origin from, e.g., strain, breaks the inversion and nonsymmorphic symmetries.

As mentioned above,
the coupling between magnetic moment and the magnetic field induce the energy shift (drift in the motion) $\delta E_{B}=-{\pmb \mu}\cdot{\bf B}$,
where ${\pmb \mu}=\hbar v_{F}e\cdot{\bf e_{k}}/(2\varepsilon)\times \hat{s}$ is the magnetic effective moment.
However, for the electromagnetic field, the induced-energy-shift is
\begin{equation} 
\begin{aligned}
\delta E=\left[\delta E_{B}-\frac{e\hbar^{2}v_{F}^{2}}{2\varepsilon(\varepsilon+\hbar v_{F}{\bf k})}{\bf E}\times{\bf k}\right]\cdot \hat{s},
\end{aligned}
\end{equation}
%{Topological spin transport of a relativistic electron}
%{Semiclassical Limit of the Dirac Equation and Spin Precession}
%{Enlarged Galilean symmetry of anyons and the Hall effect}
which is similar to the BMT equation about the spin precession in the semiclassical limit\cite{Spohn H},
and here the non-relativity correction factor is $1/(\sqrt{1-(v_{F}/c)^{2}})\approx 1$.
The above equation describes the motion of particle in the U(1) electromagnetic gauge field $({\bf B}
+\frac{\hbar^{2}v_{F}^{2}}{\varepsilon+\hbar v_{F}{\bf k}}{\bf E}\times{\bf k})$.
%{Nonadiabatic Noncyclic Geometric Phase and Ensemble Average Spectrum of Conductance in Disordered Mesoscopic Rings with Spin-Orbit Coupling}
We can see that the shift (or spin precession) $\delta E\sim 1/k^{2}(1/\varepsilon^{2})$.
That similar to the expression of the precession of the tranported spin, which can be reads 
\begin{equation} 
\begin{aligned}
\partial_{t}{\bf s}=\delta E\times {\bf s}.
%{Topological spin transport of a relativistic electron}
\end{aligned}
\end{equation}
The shift of momentum in the armchair direction of Brillouin zone can also be induced simply by the in-plane magnetic field
\cite{Dominguez F}
which breaks the reflection symmetry (reflection operator in armchair direction is $\mathcal{R}=\tau_{x}s_{y}$)
and the time-reversal-invariance (the time-reversal operator which is antiunitary is $\Theta=is_{y}\mathcal{K}$ 
where $\mathcal{K}$ denotes the complex conjugation while $is_{y}$ is the mirror operator).
%{Light-modulated 0-π transition in a silicene-based Josephson junction}
That implies that the effect of the in-plane magnetic field is similar to the in-plane exchange field (ferromagnetic or antiferromagnetic),
%{Spin-polarized current induced by a local exchange field in a silicene nanoribbon}
%{Giant magnetoresistance and perfect spin filter in silicene, germanene, and stanene}
which is effective in breaking the topologically protected gapless edge model
and thus rise the nonzero orbital magnetic moment (also the orbital effect) and the Berry curvature.
%{Testing Topological Protection of Edge States in Hexagonal Quantum Spin Hall Candidate Materials}
Both the gap under the in-plane exchange field $M$ and the perpendicular magnetic field $B_{z}$ exhibit linear relations:
for in-plane exchange field, gap $\Delta\approx 2eV/(M)eV$, while for $B_{z}$ it's $\Delta\approx 0.2 meV/T$\cite{Dominguez F}.
Thus the effect of gap opening for magnetic field is much lower than the in-plane exchange field.

The single-component Landau gauge cause a conventional circular orbits in the presence of Lorentz-like force.
However, if there is a hamonic trap potential in the presence of symmetry gauge (with two components like ${\bf A}=(B_{z}y,-B_{z}x,0)$
which lose the translation invariance in both the $x$- and $y$-direction),
%{Dynamics of neutral atoms in artificial magnetic field} 
the cyclotron orbits may becomes more complex with variant curvatures especially for the electron gas which has larger diffusion coefficient.
For the bilayer silicene (or bilayer MoS$_{2}$),
the nonzero effective mass $m^{*}$ (related to the interlayer hopping) is also shifted by the magnetic field,
as $m^{*}\rightarrow m^{*}+e{\bf B}\frac{\kappa}{m^{*}}$\cite{Horváthy P A},
where $\kappa$ is an exotic parameter in Lie algebra\cite{Grigore D R}.
Here we note that, the $\kappa$ is inversely proportional to the frequency of external harmonic trap potential if it exist,
i.e., $\kappa\propto 1/\omega$,
thus the manipulation of both the harmonic trap and the magnetic field as well as the initial momentum can be used to control the effective mass.
Note that here the effective mass is related to the initial mometum as long as the $m^{*}\nrightarrow \infty$\cite{Horváthy P A}.
%{Enlarged Galilean symmetry of anyons and the Hall effect}
Note that the trap here only affects the particles with mass, but invalid for the massless particle no matter what the frequency it is.
In fact, even without the harmonic trap, the magnetic field itself also has a coupling effect by suppressing the diffusion of the wave package.
%{Dynamics of neutral atoms in artificial magnetic field}
A singular configuration emerges at ${\bf B}=-\frac{(m^{*})^{2}}{e\kappa}$ (for gyromagnetic ratio $g\approx 2$),
where the effective mass is zero and thus give rise the Hall motion 
%{Enlarged Galilean symmetry of anyons and the Hall effect}
$\frac{1}{\hbar}\partial_{ki}\varepsilon=\varepsilon_{i\mu\nu}\frac{E_{\mu\nu}}{B}$ in the presence of disorder come from the electric field and magnetic field
(we only consider the part which affects the band structure).
%\cite{Horváthy P A}.
Since for zero effective mass under such critical magnetic field,
the particle is restricted in the lowest Landau level,
thus its quantized cyclotron orbit radius equals to the magnetic length $\ell_{B}=\sqrt{\hbar c/|eB|}$
%{Spontaneous interlayer coherence in double-layer quantum Hall systems: Charged vortices and Kosterlitz-Thouless phase transitions}
for first-order Bessel function: The angular quantum number for the first-order Bessel function\cite{Zhou X} is $l=1/2$,
which results in the same radius $\ell_{B}=\sqrt{2l\hbar c/|eB|}=\sqrt{\hbar c/|eB|}$.
%{Berry phase and anomalous velocity of Weyl fermions and Maxwell photons}
To realize the quantum Hall effect, the exchange field is required to satisfies 
$|M_{s}|>\lambda_{SOC}(1+\frac{a^{2}\lambda^{2}_{R_{2}}}{\hbar^{2}v_{F}^{2}})$,
and the band gap is generated by the anti-crossing edge models in the K-point (see Fig.3(b)),
where with the nonzero Chern number $\mathcal{C}=2$ but the spin- or valley-Chern number are all zero.
%(see one of our review paper\cite{Wu C H2}).
Since the Chern number reads 
\begin{equation} 
\begin{aligned}
\mathcal{C}=\frac{1}{2\pi}\int_{BZ}d^{2}k\Omega({\bf k}),
\end{aligned}
\end{equation}
contributed by the Pontryagin number\cite{Perez-Piskunow P M},
%{Topological phase transition and electrically tunable diamagnetism in silicene}
%{Hierarchy of Floquet gaps and edge states for driven honeycomb lattices}
we obtain $\int_{BZ}d^{2}k\Omega({\bf k})=4\pi$ for the quantum anomalous Hall phase here.
The above integral arounds the BZ is approximation result\cite{Perez-Piskunow P M} of the Chern number,
and the accuracy lifted with the decrase of Dirac-mass,
since the Berry curvature turns to a $\delta$-function at Dirac-cone in the zero-Dirac-mass-limit as we mentioned above.
%{Hierarchy of Floquet gaps and edge states for driven honeycomb lattices}

Note that here the nonzero Chern number origins from the Skyrmion spin texture where the spin rotated by $\lambda_{R_{2}}$ and 
generating the nonzero Berry curvature\cite{Ezawa M2}.
For $E_{\perp}=0$ (thus $\lambda_{R_{1}}=0$),
the band gap for quantum Hall phase is 
$\Delta=\frac{a\lambda_{R_{2}}}{\hbar v_{F}}\sqrt{\frac{\hbar^{2}v_{F}^{2}M^{2}}{\hbar^{2}v_{F}^{2}+a^{2}\lambda^{2}_{R_{2}}}-\lambda_{SOC}^{2}}$
(for $\eta\tau_{z}=-1$),
%{Valley-Polarized Metals and Quantum Anomalous Hall Effect in Silicene}
while the radius expended by the wave vector ${\bf k}_{R}$ in momentum space reads\cite{Ezawa M2}
%{Valley-Polarized Quantum Anomalous Hall Effect in Silicene}
\begin{equation} 
\begin{aligned}
{\bf k}_{R}=\frac{\sqrt{\hbar^{4}v_{F}^{4}(M_{s}^{2}-\lambda_{SOC}^{2})-a^{2}\lambda_{R_{2}}^{2}\lambda_{SOC}^{2}(2\hbar^{2}v_{F}^{2}+a^{2}\lambda^{2}_{R_{2}})}}
{\hbar v_{F}(\hbar^{2}v_{F}^{2}+a^{2}\lambda^{2}_{R_{2}})},
\end{aligned}
\end{equation}
as labeled in the last panel of Fig.1(c) for the case of $M_{s}=M_{c}=7.8$ meV
where can easily see the quantum anomalous Hall phase.
The orbital magnetic moment of the electron moves along the loop with radiu ${\bf k}_{R}$ at the lowest energy 
can be analytically obtained as
$m({\bf k})={\bf k}_{R}ev_{F}/2$.
%{Photoinduced quantum spin and valley Hall effects, and orbital magnetization in monolayer MoS2}
The Fig.1(d) shows the evolution of the band gap at K valley with different $M_{s}$ and $M_{c}$.
For the case of $\lambda_{R_{2}}=0$,
the gap vanish and the above radius reduce to the conventional form 
\begin{equation} 
\begin{aligned}
{\bf k}^{*}_{R}=\frac{\sqrt{M^{2}_{s}-\lambda_{SOC}^{2}}}{\hbar v_{F}},
\end{aligned}
\end{equation}
and it's similar to the form of Ref.\cite{Zhou X} which is for the zero-trap and zero-exchange-field case with nonzero particle mass.
%{Unconventional states of bosons with the synthetic spin–orbit coupling}
In Fig.3, we show the gap $\Delta$ and the radius ${\bf k}_{R}$ as a function of the electric field and exchange field where we set 
$M_{s}=M_{c}$.
From Fig.3(c), when $M_{s}=M_{c}=2\lambda_{SOC}=0.0078$ eV,
the radius in quantum anomalous Hall phase is about ${\bf k}_{R}=0.0012$,
and ${\bf k}_{R}$ vanish at $E_{\perp}=\frac{4\lambda_{SOC}}{\overline{\Delta}}\approx 0.0339$ eV.

The anomalous velocity induced by the Berry curvature is
\begin{equation}
\begin{aligned}
%{\bf v}_{Ai}({\bf k})=-2{\rm Im}[\varepsilon_{i\mu\nu}\partial_{t\mu}\langle\psi|\partial_{k\nu}\psi\rangle]=-\partial_{t}{\bf k}\times\Omega({\bf k}).
{\bf v}_{A}({\bf k})=i(\langle\partial_{t}\partial_{k}\psi|\partial_{k}\psi\rangle-\langle\partial_{k}\partial_{t}\psi|\partial_{t}\psi\rangle)
=-2{\rm Im}[\partial_{t}\langle\psi|\partial_{k}\psi\rangle],
\end{aligned}
%{Anomalous Bloch oscillations in one-dimensional parity-breaking periodic potentials}
%{Mapping the Berry curvature from semiclassical dynamics in optical lattices}
\end{equation}
where ${\bf k}$ is the quasimomentum of the center-of-mass,
while for the two-dimension system or the higher one, the anomalous velocity owns the Lorentz force term as
\begin{equation}
\begin{aligned}
%{\bf v}_{Ai}({\bf k})=-2{\rm Im}[\varepsilon_{i\mu\nu}\partial_{t\mu}\langle\psi|\partial_{k\nu}\psi\rangle]=-\partial_{t}{\bf k}\times\Omega({\bf k}).
{\bf v}_{A}({\bf k})=
-\partial_{t}{\bf k}\times\Omega({\bf k})
\end{aligned}
\end{equation}
and it's related to the effective force ${\bf F}$ by
\begin{equation} 
\begin{aligned}
\partial_{t}{\bf k}={\bf F}/\hbar &=\partial_{r}V({\bf r})-\frac{e}{\hbar}{\bf E}({\bf k})-\frac{e}{\hbar}{\bf v}_{g}\times{\bf B}({\bf k})\\
&=\partial_{r}V({\bf r})+\frac{e}{\hbar}\partial_{\mu}\Phi({\bf k})e^{\mu}-\frac{e}{\hbar}{\bf v}_{g}\times{\bf B}({\bf k}),
\end{aligned}
\end{equation}
%{Berry phase and anomalous velocity of Weyl fermions and Maxwell photons}
%{Berry phase correction to electron density in solids and exotic dynamics}
%{Topological spin transport of a relativistic electron}
%{Electromagnetic response of quantum Hall systems in dimensions five and six and beyond}
%{Chiral anomaly enhancement and photoirradiation effects in multiband touching fermion systems}
where $\Phi({\bf k})$ is the scalar potential of the electric field and ${\bf B}({\bf k})$ is the vector potential.
Both the scalar and vertor potential are incorporated in the Bloch wave function $\psi$.
Note that Bloch oscillation emerges when the effective force is constant,
and it's largely affected by the anomalous effect in the presence of perturbations.
Here we note that the effective Hamiltonian here contains the position-dependent period-potential term,
$H=\varepsilon({\bf k})+V({\bf r})$,
where $\varepsilon=\frac{{\bf k}^{2}(\partial_{t}{\bf k})^{2}}{2\varepsilon}|_{{\bf k}\in BZ}$ is independent of the position.
%and $\delta \varepsilon $ is the part which dependents on the position.
%Then the above equation becomes $\hbar\partial_{t}{\bf k}={\bf F}=\partial_{r}\varepsilon-e{\bf E}({\bf k})-e{\bf v}_{g}\times{\bf B}({\bf k})$.

Here the Levi-Civit{\`a} tensor is missing which distincted from the multidimensional case with Einstein summation.
%{Enlarged Galilean symmetry of anyons and the Hall effect} 
${\bf v}_{g}=\frac{\partial H}{\hbar\partial {\bf k}}+{\bf v}_{A}$ is group velocity (after Berry correction) of the center of the wave package,
where $H=\varepsilon({\bf k})+V({\bf r})$ is the summation of the quasienergy and the spin-independent perturbation potential.
Through the above Hall motion, the group velocity can be rewritten as 
\begin{equation} 
\begin{aligned}
{\bf v}_{g}=
\frac{\hbar}{e{\bf B}}\partial_{t}{\bf k}+\frac{{\bf E}}{{\bf B}},
\end{aligned}
\end{equation}
thus Berry curvature here vanish for ${\bf B}=0$.
Through the form of expression of the $\hbar\partial_{t}{\bf k}$,
we can see that the motion of electrons is along the applied fields, i.e., along the $k_{z}$-direction.
While in the case of without the perturbations (without electric field or magnetic field),
the charge density $\rho$ at the band crossing point obeys the continuity equation
\begin{equation} 
\begin{aligned}
\partial_{t}\rho+\nabla_{{\bf k}}\cdot(\rho\hbar\partial_{t}{\bf k})=0
\end{aligned}
\end{equation}
in quasimomentum space,
%{Berry phase correction to electron density in solids and exotic dynamics}
and that's also valid for the multiband touching models including the bilayer silicene or Weyl semimetal\cite{Bradlyn B,Ezawa M}.
The above continuity equation is also valid in phase space (including the coordinate and quasimomentum).
But that needs the distribution function to satidfies $\partial_{t}f=0$,
or in the scattering form of Boltzmann-Vlasov equation\cite{Liu Y,Chang M C}:
$\partial_{t}f+{\bf v}_{g}f-\frac{e}{\hbar}\cdot\nabla_{{\bf k}}=(\partial_{t}f)_{collision}=0$,
%{Effects of the relative motion of different particles on the wave instability in dusty plasmas}
which is satisfied when the scattering broadening is smaller than the bandwidth.
Through the above corrected group velocity, the interband transition can be described by the momentum operator
\begin{equation} 
\begin{aligned}
\mathcal{P}^{\eta}=m_{0}&\left[\langle \psi|\frac{\partial H}{\hbar\partial {\bf k}}|\psi' \rangle +\hbar\partial_{t}{\bf k}\cdot\Omega({\bf k})\right]\\
=m_{0}&\left[ (v_{F}+\frac{i\eta a\lambda_{R_{2}}}{\hbar})(1+\eta\frac{m_{D}^{\eta s_{z}\tau_{z}}}{\varepsilon})\right.\\
&\left.+[\partial_{r}V({\bf r})+\frac{e}{\hbar}\partial_{\mu}\Phi({\bf k})e^{\mu}-\frac{e}{\hbar}{\bf v}_{g}\times{\bf B}({\bf k})]
\cdot\frac{1}{2 }\frac{2\eta \hbar^{2}v_{F}^{2}m^{\eta s_{z}\tau_{z}}_{D}}{2\varepsilon(4(m^{\eta s_{z}\tau_{z}}_{D})^{2}+\hbar^{2}v_{F}^{2}k^{2})}.\right],
\end{aligned}
\end{equation}
%{Spin-valley optical selection rule and strong circular dichroism in silicene}
%{Coupled Spin and Valley Physics in Monolayers of MoS2 and Other Group-VI Dichalcogenides}
where $m_{0}$ is the free electron mass.
The plot of $\mathcal{P}^{\eta}$ for $\eta=1$ (K valley) is presented in Fig.4.
During the interband transition, the kinetic term won't induce the spin flip while the Rashba-coupling $\lambda_{R_{2}}$
will induce the spin flip in a topological insulator\cite{Ezawa M3},
which mays the spin $s_{z}$ no more a good quantum number.

It's obvious that the symmetry broken (or parity broken) together with the time-dependence (of band) give rise the anomalous velocity
and the non-adiabatic correction.
While the adiabatic approximation requires the large band gap $m_{D}^{\eta s_{z}\tau_{z}}>L{\bf F}/2$ 
(to prevent the excited particles through the gap)
in the absence of the scattering and perturbations.
%{Anomalous Bloch oscillations in one-dimensional parity-breaking periodic potentials}
%{Berry Phase, Hyperorbits, and the Hofstadter Spectrum}
The anomalous velocity is vanishes for the moment-free spin(or pseudospin)-1 massless particle which can easily be found in the Maxwell metal.
%{Berry phase and anomalous velocity of Weyl fermions and Maxwell photons}
%{Topological Maxwell Metal Bands in a Superconducting Qutrit}
For such a particle, the above effective Hamiltonian matrices (4)-(6) are still valid 
just by replacing $\hbar v_{F}(k_{x}\pm ik_{y})$ with $\frac{1}{\sqrt{2}}\hbar v_{F}(k_{x}\pm ik_{y})$.
%{Chiral anomaly enhancement and photoirradiation effects in multiband touching fermion systems}
The quasienergy here is with a period of $2\pi/L$ in momentum space which is estimated as 1 here,
thus in first Brillouin zone it has $\varepsilon(k)=\varepsilon(k+2\pi/L)$.
%{Electrical tuning of valley magnetic moment through symmetry control in bilayer MoS2}
The density of monopole charge (which carried by the monopole at the origin of the momentum space) 
%{Topological spin transport of a relativistic electron}
is nonzero for the Bloch band only when this Bloch band degenerate with other bands\cite{Murakami S},
and it can be evaluated as $\frac{1}{2\pi}\int_{BZ}d{\bf k}\partial_{k}\Omega({\bf k})=2\hat{s}$.
We see that it's different from the above mentioned case,
since the Bloch band degenerate with other bands in momentum space and the Berry curvature is no more a curl.
%{Berry phase correction to electron density in solids and exotic dynamics}
%{Beyond Dirac and Weyl fermions: Unconventional quasiparticles in conventional crystals}
%{Phase transition between the quantum spin Hall and insulator phases in 3D: emergence of a topological gapless phase}
In this case, the Berry phase is variant with the perturbation-dependent phase factor.

The intrinsic quantum spin Hall conductivity due to the anomalous velocity and electron trajectories reads\cite{Tahir M}
\begin{equation} 
\begin{aligned}
\sigma_{xy}^{s}=\frac{e^{2}}{\hbar}\int_{BZ}\frac{d^{2}k}{4\pi^{2}}[f_{s_{z}=1}-f_{s_{z}=-1}]\Omega({\bf k}).
\end{aligned}
\end{equation}
For the case Fermi-level within the band gap, the spin Hall conductivity reads
\begin{equation} 
\begin{aligned}
\sigma_{xy}^{s}=\frac{e^{2}}{4h}\sum_{\eta\tau_{z}}({\rm sgn}[m_{D}^{\eta\tau_{z},s_{z}=1}]-{\rm sgn}[m_{D}^{\eta\tau_{z},s_{z}=-1}]).
\end{aligned}
\end{equation}
According to the Dirac-mass mentioned above, the requirements for the nonzero spin Hall conductivty can be obatined as
\begin{equation} 
\begin{aligned}
\eta\lambda_{SOC}\tau_{z}+M>\frac{\overline{\Delta}}{2}E_{\perp}\tau_{z}+\eta\hbar v_{F}^{2}\mathcal{A}^{2}/\Omega,\\
{\rm and}\ \frac{\overline{\Delta}}{2}E_{\perp}\tau_{z}+\eta\hbar v_{F}^{2}\mathcal{A}^{2}/\Omega<0,
\end{aligned}
\end{equation}
or
\begin{equation} 
\begin{aligned}
\eta\lambda_{SOC}\tau_{z}+M<\frac{\overline{\Delta}}{2}E_{\perp}\tau_{z}+\eta\hbar v_{F}^{2}\mathcal{A}^{2}/\Omega,\\
{\rm and}\ \frac{\overline{\Delta}}{2}E_{\perp}\tau_{z}+\eta\hbar v_{F}^{2}\mathcal{A}^{2}/\Omega>0.
\end{aligned}
\end{equation}
For the case Fermi level lies within the conduction band,
the spin Hall conductivity can be obtained as
\begin{equation} 
\begin{aligned}
\sigma_{xy}^{s}=\frac{e^{2}}{4h}\sum_{\eta\tau_{z}}
[\frac{m_{D}^{\eta\tau_{z},s_{z}=1}}{\varepsilon_{s_{z=1}}}-\frac{m_{D}^{\eta\tau_{z},s_{z}=-1}}{\varepsilon_{s_{z=-1}}}],
\end{aligned}
\end{equation}
%{Dc and ac transport in silicene}
Our above analytical calculations are also agree with the results of Ref.\cite{Vargiamidis V}.
While for the intrinsic quantum valley Hall conductivity,
it can be obtained through the similar procedure but focus on the differece between the cases of $\eta=1$ and $\eta=-1$.
Related to the Boltzman equation in finite temperature but in the absence of vertex correction between different tiles with different self-energy, 
%{Spatial Correlations and the Insulating Phase of the High- Cuprates Insights from a Configuration-Interaction-Based Solver for Dynamical Mean Field Theory support}
the above Hall conductivity in quantum anomalous Hall phase can be rewritten as
%{Intrinsic Versus Extrinsic Anomalous Hall Effect in Ferromagnets}
\begin{equation} 
\begin{aligned}
\sigma_{xy}=\frac{e^{2}}{2\hbar}\int_{BZ}\frac{d^{2}k}{4\pi^{2}}\sum_{\psi\psi'}(\varepsilon(\psi)-\varepsilon(\psi'))^{-1}\partial_{\varepsilon}f(\varepsilon(\psi))
\times\Omega({\bf k}),
\end{aligned}
\end{equation}
where $f(\varepsilon(\psi))$ is the Dirac-Fermi distribution function.

\section{Perturbation from the impurity scattering potential and periodic off-resonance light}

The impurity is important to the intervalley scattering,
especially in the bulk part where the valley mixing is missing\cite{Xiao D2}.
%{Valley filter and valley valve in graphene}
In the presence of nonzero impurity scattering angle with a single nonmagnetic (without the coupling with spin) impurity, 
i.e., the position-dependent Gaussian scattering potential in the $T$-matrix approximation\cite{Onoda S},
the density of state (DOS) reads
\begin{equation} 
\begin{aligned}
D({\bf k},\omega)&=\int_{BZ}\frac{d^{2}k}{(2\pi)^{2}}(f_{\varepsilon_{m}}-f_{\varepsilon_{n}})\\
&{\rm Im}
[G_{{\bf k}}(\varepsilon_{m}-\varepsilon_{n})T(\omega)G_{{\bf k}}
(\varepsilon_{m}-\varepsilon_{n}+\Delta{\bf k})-G_{{\bf k}}(\varepsilon_{n}-\varepsilon_{m})T(\omega)G_{{\bf k}}(\varepsilon_{n}-\varepsilon_{m}+\Delta{\bf k})],
\end{aligned}
\end{equation}
%The expression of the DOS is distinct from the conductivity since it takes the imaginary part of the lattice Green's function in the
%defect configuration.
%{Optical Absorption in Non-Crystalline Semiconductors}
Here the effect of $T(\omega)$ is similar to the vertex function except that the vertex function is a connection between different frequencies
but with the same momentum while the $T(\omega)$ here is a connection between different momentums which is related to the scattered wave vector
but with the same frequency\cite{Wu C H1}.
%{Quasiparticle scattering and local density of states in graphite}
%{Quasiparticle scattering and local density of states in the d-density-wave phase}
%{Quasiparticle scattering interference in high-temperature superconductors}
%{Spatial Correlations and the Insulating Phase of the High- Cuprates Insights from a 
%Configuration-Interaction-Based Solver for Dynamical Mean Field Theory}
The direction of $T(\omega)$ is perpendicular to the boundary between the two distinguish momentum tiles,
and weighted by the (momentum-resolved) DOS or the spectral function,
%\cite{Haule K,Go A,Xu W}
%{Photoinduced Topological Phase Transition and a Single Dirac-Cone State in Silicene}
which is also related to the low-energy behaviours of 
optical conductivity and the Hall conductivity.
%{Spatial Correlations and the Insulating Phase of the High- Cuprates Insights from a Configuration-Interaction-Based Solver for Dynamical Mean Field Theory}
%{Hidden Fermi liquid_ self-consistent theory for the normal state of high-T c superconductors}
%{Coupled charge and spin dynamics in a photo-excited Mott insulator}
%{Pseudogaps in an Incoherent Metal}
%{Hole spectral functions in lightly doped quantum antiferromagnets}
%{Hidden Fermi liquid, scattering rate saturation, and Nernst effect a dynamical mean-field theory perspective
%{Real-Time Evolution Using the Density Matrix Renormalization Group}
For the nonmagnetic impurity
(or the weak magnetic ordering impurities like W- or Mo-silicene), 
%{Silicene and transition metal based materials prediction of a two-dimensional piezomagnet
the $T(\omega)$ has
%\cite{Wang Q H}
\begin{equation} 
\begin{aligned}
T^{-1}(\omega)=\frac{1}{V_{s}\sigma_{z}}-\frac{1}{\hbar^{2}}\int_{BZ}\frac{d^{2}k}{4\pi^{2}}G_{{\bf k}}(\varepsilon_{m}-\varepsilon_{n}),
\end{aligned}
\end{equation}
%{Intrinsic Versus Extrinsic Anomalous Hall Effect in Ferromagnets}
%{Quasiparticle scattering and local density of states in graphite}
%{Quasiparticle scattering and local density of states in the d-density-wave phase}
%{Quasiparticle scattering interference in high-temperature superconductors}
where $V_{s}$ is the single scalar impurity scattering potential with the $z$-direction spin-polarization
(here we note that the conductance in the presence of scalar impurity is larger than that in the presence of magnetic impurity).
%{Testing Topological Protection of Edge States in Hexagonal Quantum Spin Hall Candidate Materials}
The impurity scattering potential here is assumed to be momentum-independent but energy (frequency)-dependent,
and thus it is in a Gaussian scattering form with a $\delta$-function: $\delta({\bf r}-{\bf r}_{{\rm imp}})$.
The Keldysh Green's function here $G_{{\bf k}}(E_{m}-E_{n})$ independent of the scattering,
and it reads\cite{Onoda S}
\begin{equation} 
\begin{aligned}
G_{{\bf k}}(\varepsilon_{m}-\varepsilon_{n})=(\varepsilon-H-n_{{\rm imp}}T(\omega))^{-1},
\end{aligned}
\end{equation}
where $n_{{\rm imp}}T(\omega)$ is the scattering-independent self-energy term.
The perturbated Hamiltonian under the impurity scattering potential is
\begin{equation} 
\begin{aligned}
H_{V}=H_{0}+V\frac{1}{\varepsilon-H_{0}}V+V\frac{1}{\varepsilon-H_{0}}V\frac{1}{\varepsilon-H_{0}}V+O((\varepsilon-H_{0})^{-3}).
\end{aligned}
\end{equation}
Note that here the nonmagnetic impurity scattering potential is assumed independent of momentum and thus has a $\delta$-function singularity,
%{Quasiparticle scattering and local density of states in graphite}
%{Quasiparticle scattering and local density of states in the d-density-wave phase}
%{berry's phase and absence of back scattering in carbon nanotubes}

According to the second-order effect (which is unique to the off-resonance light) of the effective photocoupling process,
where the quasienergy is shifted by absorbing (emitting) a certain quantity photons,
as $\varepsilon\pm \hbar\Omega/2$, i.e., it's shifted by a half-integer multiples of $\hbar \Omega/2$.
%{Hierarchy of Floquet gaps and edge states for driven honeycomb lattices}
%{Photoinduced Topological Phase Transition and a Single Dirac-Cone State in Silicene}
The transport properties during such process can be detected by the Green's function with complex variable (quasienergy),
and the decimation method\cite{Pastawski H M}.
For the perturbation from the time-dependent periodic off-resonance light $V(T+t)=V(t)$,
the monochromatic-perturbated Hamiltonian can be obtained by the simple Schr\"{o}dinger equation $H_{V}\phi=E\phi$ using the Floquet technique,
with Floquet Hamiltonian in a tridiagonal form as
\begin{equation}
\begin{aligned}
H_{V}=
\begin{pmatrix}
\ddots\\
\cdot\cdot\cdot&V_{-N} &H_{-N}&V_{+N}&0&0&\cdot\cdot\cdot\\ 
\cdot\cdot\cdot&0 &V_{-N}&H_{0}&V_{+N}&0\cdot\cdot\cdot\\ 
\cdot\cdot\cdot&0 &0&V_{-N}&H_{+N}&V_{+N}&\cdot\cdot\cdot\\ 
 & & & & & &\ddots\\
\end{pmatrix},
\end{aligned}
\end{equation}
and 
\begin{equation}
\begin{aligned}
\phi=(\cdot\cdot\cdot,\phi_{-N},\phi_{0},\phi_{+N},\cdot\cdot\cdot)^{T}
\end{aligned}
\end{equation}
thus the $H_{V}$ can be obtained as
\begin{equation}
\begin{aligned}
H_{V}=&H_{0}+V_{-N}\frac{1}{\varepsilon-H_{-N}}V_{+N}+V_{+N}\frac{1}{\varepsilon-H_{+N}}V_{-N}+O(V_{\pm N}^{2})\\
\approx & H_{0}+\frac{[H_{-1},H_{+1}]}{\hbar\Omega}+O(\frac{\mathcal{A}^{4}}{\Omega^{2}}),
\end{aligned}
\end{equation}
with the interaction term $V_{N}=\frac{1}{T}\int^{T}_{0}H(t)e^{-iN\hbar\omega t}$ where $H(t)$ is the one appear in Appendix.A.
$H_{0}$ can be evaluated as the one-period mean value of $H(t)$, i.e., $H_{0}=\frac{1}{T}\int^{T}_{0}H(t)dt$.
The above perturbated effective Hamiltonian can be related to the evolution operator with the effect of Berry curvature by\cite{Kitagawa T}
\begin{equation}
\begin{aligned}
H_{V}=\frac{i\hbar}{T}{\rm log}U,
\end{aligned}
\end{equation}
with 
\begin{equation}
\begin{aligned}
U=e^{-iTH_{0}}\mathcal{P}e^{i\int_{C}A({\bf k})d{\bf k}},
\end{aligned}
\end{equation}
where $\mathcal{P}$ is the path operator of the electron contour $C$.
%{Hierarchy of Floquet gaps and edge states for driven honeycomb lattices}
%{Topological spin transport of a relativistic electron}
%{Photoinduced quantum spin and valley Hall effects, and orbital magnetization in monolayer MoS2}
%{Transport properties of nonequilibrium systems under the application of light Photoinduced quantum Hall insulators without Landau levels}
%{Photoinduced Topological Phase Transition and a Single Dirac-Cone State in Silicene}

While for the five-diagonal one, 
we can using the follwing equation with real space renormalization group 
\begin{equation}
\begin{aligned}
\begin{pmatrix}
H_{-N}&V_{+N}&V_{+N+1}\\ 
V_{-N}&H_{0}&V_{+N}\\ 
V_{-N-1}&V_{-N}&H_{+N}\\ 
\end{pmatrix}
\begin{pmatrix}
\phi_{-N}\\ 
\phi_{0}\\ 
\phi_{+N}\\ 
\end{pmatrix}
=\varepsilon
\begin{pmatrix}
\phi_{-N}\\ 
\phi_{0}\\ 
\phi_{+N}\\ 
\end{pmatrix},
\end{aligned}
\end{equation}
with the Fourier index $N$ and the Bloch wave function in zeroth Fourier model is
\begin{equation}
\begin{aligned}
\phi_{0}=\frac{V_{-N}}{\varepsilon-H_{0}}\phi_{-N}+\frac{V_{+N}}{\varepsilon-H_{0}}\phi_{+N}.
\end{aligned}
\end{equation}
%{Tight Binding’methods in quantum transport through molecules and small devices: From the coherent to the decoherent description.}

The circular polarized light-induced periodically-driven nonequilibrium system results in a dc-driven charge current,
which disobey the current continuity $\nabla \cdot{\bf J}+\frac{\partial \rho}{\partial t}=0$ and 
the probability current conserved $-\frac{i\hbar}{2m}(\phi^{*}\nabla\phi-\phi\nabla\phi^{*})$ and
the Gaussian distribution (wigner-Dyson type), which 
can be represented by a variant reservior response form 
in a determined Landau level with the frequency $\omega$ before the optical coupling:\cite{Kitagawa T}
\begin{equation} 
\begin{aligned}
{\bf J}_{{\rm res}}=&\sum_{n,b}\int 
\frac{d\omega}{2\pi}t_{a}^{2}\rho_{a}(\omega+n \hbar\omega_{l})t_{b}^{2}\rho_{b}(\omega)G_{ij}(n,\omega)
(f_{b}(\omega)-f_{a}(\omega+n \hbar\omega_{l})),\\
G_{ij}(n,\omega)=&\int dt e^{in\hbar\omega_{l}t}\int dt'G_{ij}^{{\rm ret}}(t,t')e^{i(\omega+i0^{+})(t-t')},\\
\rho_{a}(\omega+n \hbar\omega_{l})=&\frac{1}{N}\sum_{{\bf k}}\delta(\omega+n \hbar\omega_{l}-t|\epsilon_{{\bf k}}|),\\
\rho_{b}(\omega)=&\frac{1}{N}\sum_{{\bf k}}\delta(\omega-t|\epsilon_{{\bf k}}|),\ t=\frac{2\sqrt{3}\hbar v_{F}}{3a}\approx 1.6 eV,\ v_{F}\approx 5.5\times 10^{5}\ {\rm m/s}\\
f_{b}(\omega)=&\frac{1}{1+{\rm exp}[\beta_{b}(\omega-\mu_{b})]},\\
f_{a}(\omega+n \hbar\omega_{l})=&\frac{1}{1+{\rm exp}[\beta_{a}(\omega+n \hbar\omega_{l}-\mu_{a})]},\\
\end{aligned}
\end{equation}
%{v_{F}: Protected edge states in silicene antidots and dots in magnetic field}
where the $a$ corresponds the channel which coupling with the photons (absorbs (or emits) $n$ photons),
while the $b$ is the one which not couple with the photons
(i.e., describe the transport between two leads $a$ and $b$).
$f_{a/b}$ is the Fermi-Dirac distribusion function, 
$\omega_{l}$ denotes the frequency of the light,
$\rho_{a/b}$ is DOS per unit cell of each channel,
and $G_{ij}^{{\rm ret}}(t,t')$ is the retarded Green's function
$G_{ij}^{{\rm ret}}(t,t')=-i\theta(t-t')\langle\{c_{i}(t),c_{j}^{\dag}(t)\}\rangle$ ($\theta(t-t')$ is the Heaviside step function),
and th term $G_{ij}^{{\rm ret}}(t,t')e^{i(\omega+i0^{+})(t-t')}$ 
can be replaced by the advanced Green's function as $G_{ij}^{{\rm adv}}(t,t')e^{i(\omega+i0^{-})(t'-t)}$ 
with $G_{ij}^{{\rm adv}}(t,t')=i\theta(t'-t)\langle\{c_{i}(t),c_{j}^{\dag}(t)\}\rangle$ in the above equation
since the relation $G_{ij}^{{\rm ret}}(t,t')=(G_{ij}^{{\rm adv}}(t,t'))^{*}$.
%{Interaction quench in the Hubbard model_ Relaxation of the spectral function and the optical conductivity}
While for the Matsubara frequency $\omega_{M}$, wejust need to replace the real time by the imaginary time $\tau$.
The reservior variables can be well described by the master equation in the Liouville space
with the unperturbed density operator $\mathcal{J}$
\begin{equation}   
\begin{aligned}
\partial_{t}\mathcal{J}=-i[H,\mathcal{J}]+\mathcal{K}\sum_{i}[O_{i}\mathcal{J}O_{i}^{\dag}-\frac{1}{2}(O_{i}^{\dag}O_{i}\mathcal{J}+\mathcal{J}O_{i}^{\dag}O_{i})]\equiv\mathcal{L}\mathcal{J},
\end{aligned}
\end{equation}
where $\mathcal{J}$ corresponds to the pure state or mixed state and $O_{i}$ is the Lindblad operator describing the bath coupling.
%For a detail discussion about the matter+dielectric mediated+reservoir system, see Ref.\cite{Behunin R O}.

The diagonal Floquet Green's function
\begin{equation}   
\begin{aligned}
G_{{\rm dia}}^{N}=\frac{1}{\varepsilon-(H_{-1}+V_{+1}\frac{1}{\varepsilon-H_{0}}V_{-1})-(V_{+N+1}+V_{+N}\frac{1}{\varepsilon-H_{0}}V_{+N})\frac{1}{\varepsilon-(H_{+N}+V_{-N}\frac{1}{\varepsilon-H_{0}}V_{+N})}(V_{-N-1}+V_{-N}\frac{1}{\varepsilon-H_{0}}V_{-N})},
\end{aligned}
\end{equation}
The retarded or advanced diagonal Floquet Green's functions obeys
\begin{equation}   
\begin{aligned}
G_{{\rm dia}}^{N+1}=[\varepsilon{\bf I}-(H_{+N+1}+V_{-N}G_{{\rm dia}}^{N}V_{-N})]^{-1}.
\end{aligned}
\end{equation}
%{Tight Binding’ methods in quantum transport through molecules and small devices: From the coherent to the decoherent description.}

\section{Anomalous Andreev reflection in Josephson device under electric field and off-resonance light}

The Josephson effect\cite{Josephson B D} 
is the earliest explanation about the tunneling between two superconductors separated by a oxide layer or quantum dot\cite{Szombati D B,Pala M G}.
The tunneling strength is $S=2\pi|V|^{2}\delta(\omega-\mu_{n}-\varepsilon_{n}(m_{Dn}^{\eta\sigma_{z}\tau_{z}}=0))$,
where 
\begin{equation} 
\begin{aligned}
V=\sum_{{\bf q},s_{z}}(T_{{\bf k},({\bf k}+{\bf q})}c^{\dag}_{{\bf k},s_{z}}d_{({\bf k}+{\bf q}),s_{z}}+H.c.)
-m_{Dn}^{\eta\sigma_{z}\tau_{z}}(e^{i\phi}c^{\dag}_{{\bf k},s_{z}}d_{({\bf k}+{\bf q}),s_{z}}+H.c.)\\
\end{aligned}
\end{equation}
is the tunneling matrix element,
%{Unconventional Josephson Signatures of Majorana Bound States}
%{TWO-PARTICLE TUNNELING PROCESSES BETWEEN SUPERCONDUCTORS}
%{TUNNE LING BETWEEN SUPERCONDUCTORS  Vinay Ambegaokar and Alexis Baratoff}
where $c^{\dag}$ $d^{\dag}$ are the creation operators of single-particle state in the left and right leads, respectively.
$\omega$ is the excitation energy of single-particle state,
$\mu$ is the chemical potential in the middel region. $m_{D}^{\eta\sigma_{z}\tau_{z}}$ is the Dirac-mass (excitation gap) in the normal region
(see Appendix),
thus $\varepsilon(m_{D}^{\eta\sigma_{z}\tau_{z}}=0)$ is the quasienergy (single particle energy) in the middle region.
The many-body wave function which decrease exponentially in the normal region can be represented as
$\Psi(x)\propto e^{-x/\xi_{s}}e^{d/\ell}e^{i\phi/2}$,
where $\xi$ is the superconducting coherance length, $x$ is the mean free path in the bulk region (shorter than the middle normal region),
$\ell=0.47$ \AA\ is the characteristic length scale of silicene in the normal region (without the proximity-induced superconductivity or the magnetism), 
$d$ is the vertical distance to the nonmagnetioc impurity within the substrate,
$\phi$ is the phase difference between the left and right superconducting leads
which can be ignored if the middle region is replaced by an ordinary superconductor\cite{Jiang L}.
%difference between left and right superconducting leads,as we have discussed\cite{Wu C H7}.
Here note that we imaging the ideal interface where the Fermi wavelength in superconducting leads is much shorter than the superconducting coherence length.

For excitation gap in middle region smaller than the superconducting gap, $m_{D}^{\eta\sigma_{z}\tau_{z}}<\Delta_{s}$,
the Josephson effect emerges through the formation and disruption of Cooper pairs with the process of Andreev reflection
and the mixing between the conduction band and valence band in the interface state with superconductor.
$\Delta_{s}$ is the superconducting gap (complex pair potential) which obeys the BCS relation 
and can be estimated as $\Delta_{s}=\Delta_{0}{\rm tanh}(1.74\sqrt{T_{c}/T-1})e^{i\phi/2}$ 
(here we only consider the right superconducting lead)
%\cite{Zhou X,Annunziata G}
with $\phi$ the macroscopic phase-difference between the left and right superconducting leads,
$\Delta_{0}$ the zero-temperature energy gap which estimated as 0.001 eV here and 
%{Andreev reflection in graphene nanoribbons}
$T_{c}\approx 5.66\times 10^{-4}$ eV the superconducting critical temperature.

The perturbations including the electric field and the off-resonance light can be taken into accout within the computation of Andreev bound state levels,
it's found that, for $\lambda_{R_{1}}=\mu=m_{D}^{\eta s_{z}\tau_{z}}\ll \varepsilon$,
the anomalous Andreev reflection is dominating\cite{Beiranvand R}
during the electronic tunneling in the Josephon junction.
In the retro-reflection regime with subgap energies $\varepsilon<\Delta_{s}$,
in contrast to the conventional Andreev reflection whose backscattered hole pass through the valence band\cite{Beiranvand R}
and with a spin-flip process (due to the $\lambda_{R_{2}}$), 
the anomalous Andreev reflection happen with an another electron
with opposite spin come from the valence band and scatter a hole lies still in the conduction band.
%{Tunable anomalous Andreev reflection and triplet pairings in spin-orbit-coupled graphene}
%{Superconducting proximity effect in silicene- Spin-valley-polarized Andreev reflection nonlocal transport, and supercurrent}
Consider the $\lambda_{R_{2}}$, the anomalous Andreev reflection is still possible due to the band splitting as presented in Fig.1(d).
Another anomalous effect in the presence of $\lambda_{R_{1}}$ is due to the reversing of Berry phase\cite{Zhai X}.
Distinct from the Berry phase obtained above,
the additional phase factor induced by $\lambda_{R_{1}}$ will cause the Berry phase of monolayer silicene changes to $2\pi$ at valley K from the previou value 
$\pi$, while the Berry phase of bilayer silicene is changes to $\pi$ at valley K from the previou value $2\pi$.
Note that here the result is valid for $\lambda_{R_{1}}\gg\lambda_{SOC}$ by the modulation of electric field.
The Andreev reflection is enhanced in the monolayer silicene but reduced in bilayer silicene due to the 
transition of Berry phase,
due to the constructive interference by the $\pi$ Berry phase and destructive interference by the $2\pi$ Berry phase\cite{Aronov A G,Zhai X}.
this is similar to the effect of bias voltage in finite doped system;
the low bias voltage gives rise the retro-reflection while high bias voltage gives rise the specular one.

For the superconductor-ferromagnet-superconductor (SFS) junction,
the Cooper pairs can be described by the above many-body wave function as
$\Psi(x)\propto e^{-x/\xi_{s}}e^{d/\ell}e^{i\phi/2}e^{-x/\xi_{f}}{\rm cos}(x/\xi_{f})$,
where $\xi_{f}\propto\sqrt{D_{f}/M}$\cite{Alidoust M} is the characteristic length for ferromagnetic silicene 
with the diffusion coefficient $D_{f}\propto v_{F}\tau$ and the exchange field $M$.
%{Josephson junction through a disordered topological insulator with helical magnetization}
The quasiparticle mean free time $\tau\rightarrow \infty$ for ballistic transport.
In fact both the exchange field and the magnetic field can suppress the diffusion by coupling the motions in each diection as we mentioned above,
and that can be described by the covariant derivative as shown in Ref.\cite{Zyuzin A}.
In the regime of excitation quasienergy $\varepsilon\le \Delta_{0}$ and $\mu\approx m_{D}^{\eta s_{z}\tau_{z}}$,
it's dominated by the conventional normal reflection while the reflected particles with minority spin is much less\cite{Beiranvand R}.
In Fig.5, we show the Andreev bound state level in SNS-junction versus the phase differentunder different conditions.
The detail computation procedure is presented in the Appendix.

%\section{Results and discussion}
\section{Conclusions}

We discuss the effect of time-dependent scalar or vector potential and nonzero Berry curvature on the electronic transport in the silicene,
where the degrees of freedom (spin, pseudospin, and valley) related to the center-of-mass play an important role.
The anomalous effects due to the Berry correction rised by the time-dependent scalar or vector potential
as well as the time-dependent band structure and Bloch band states, 
like the anomalous velocity term, anomalous Bloch oscillation in the presence of an
effective force, and the anomalous Andreev reflection.
% where the reflected holes pass through the valence band.
%in the case of small Dirac-mass and chemical potential and the large excitation quasienergy.
%{Tunable anomalous Andreev reflection and triplet pairings in spin-orbit-coupled graphene}
%{Anomalous Bloch oscillations in one-dimensional parity-breaking periodic potentials}
%The perturbations from the scattering by the charged impurities and the time-dependent off-resonance light are presented in the Appendix.B.
The topological spin/valley transport as well as the momentum of center-of-mass of the wave package 
are affected by the applied electric field, magnetic field, and the off-resonance circularly polarized light.
The anomalous velocity due to the Berry curvature
(including a Lorentz-like term) shifts the electrons in the direction transverse to the electric field and magnetic field,
%{Coupled Spin and Valley Physics in Monolayers of MoS2 and Other Group-VI Dichalcogenides}
and gives rise to the spin (transverse) Hall conductivity
as we investigate in this article.
The model detected here is the silicene-like two-dimension hexagonal system,
where the motion of the Skyrmion spin texture (in quantum anomalous Hall phase) carried by the spin current is rather weak
(compared to the one in quantum spin Hall phase).

The momentum operator about the interband transition involving the Berry correction is obtained,
where the corrected velocity term is used including the effects of external periodic potential and the effective force in semicalssical dynamics.
%{Anomalous Bloch oscillations in one-dimensional parity-breaking periodic potentials}
For perturbation related to the valley index only, like the right(+)- or left(-)-handed off-resonance light, the 
momentum operator satisfies: $\mathcal{P}_{+}^{K}=\mathcal{P}_{-}^{K'}$,
i.e., the momentum operator in one valley is just the time-reversal of the another valley,
and that also valid for the selection excitation.
%{Coupled Spin and Valley Physics in Monolayers of MoS2 and Other Group-VI Dichalcogenides}
We also found that, in the presence of orbital degree of freedom and the off-resonance circularly polarized light,
the valley mixing (through the edge states), valley polarization and the orbital magnetic moment (or orbital magnetization) affected largely by the 
light field with the selection rule.
%{Coupled Spin and Valley Physics in Monolayers of MoS 2 and Other Group-VI Dichalcogenides}
%{Photoinduced quantum spin and valley Hall effects, and orbital magnetization in monolayer MoS2}
The symmetry broken due to the perturbations together with the time-dependence of band give rise the anomalous velocity
and the non-adiabatic correction,
while the adiabatic approximation requires the large band gap $m_{D}^{\eta s_{z}\tau_{z}}>L{\bf F}/2$ 
(to prevent the excited particles through the gap)
in the absence of the scattering and perturbations,
thus our result won't be valid for the single band case which with very small effective force ${\bf F}$
and thus can be treated as the adiabatic case.
%{Mapping the Berry curvature from semiclassical dynamics in optical lattices}
For the Andreev reflection in the presence of electric-field-induced Rashba-coupling,
where we imaging a ideal interface between the conductor and the superconductor leads
%{Andreev reflection and Klein tunneling in graphene}
(note that here the Andreev reflection can't be happen in the middle regime if the conductor is replaced by a insulator barrier),
the spin-flip becomes possible even during the process of backscattering,
and the anomalous equal-spin Andreev reflection\cite{Beiranvand R}
can happen in regime $\varepsilon<\mu+m_{D}^{\eta s_{z}\tau_{z}}$ in the absence of $\lambda_{R_{1}}$,
where the reflected hole lies in the conduction band through the backscattering.
The investigation of the anomalous effects induced by the Berry curvature is helpful to understanding the semiclassical dynamics,
quantum anomalous Hall effect\cite{Jungwirth T,Onoda M},
Bloch electron system, and even the condensate matter system\cite{Price H M} or magneto-electronic devices\cite{Zhang Y},
we mainly focus on the silicene-like hexagonal lattice system in low-energy Dirac tight-binding model in this article.
Our results can also be used to the silicene-like topological insulators,
like the germanene, tinene, Mo$S_{2}$, black phosphorus.

%The index "$n$" in above equations is to distinct them from the Dirac-mass and quasienergy spectrum in superconducting regions,
%which are 
%\begin{equation} 
%\begin{aligned}
%m_{D}^{\eta\sigma_{z}\tau_{z}}=&|\eta\lambda_{{\rm SOC}}s_{z}\tau_{z}+M_{s}s_{z}+M_{c}|,\\
%\varepsilon=&s\sqrt{(\sqrt{\hbar^{2}v_{F}^{2}{\bf k}^{2}+(m_{D}^{\eta\sigma_{z}\tau_{z}})^{2}}+s \mu_{s})^{2}+\Delta_{s}^{2}},
%\end{aligned}
%\end{equation}
%where $\Delta_{s}$ is the superconducting gap (complex pair potential) which obeys the BCS relation 
%and can be estimated as $\Delta_{s}=\Delta_{0}{\rm tanh}(1.74\sqrt{T_{c}/T-1})e^{i\phi/2}$ 
%(here we only consider the right superconducting lead)\cite{Zhou X,Annunziata G}
%with $\phi$ the macroscopic phase-difference between the left and right superconducting leads,
%$\Delta_{0}$ the zero-temperature energy gap which estimated as 0.001 eV\cite{Rainis D} here and $T_{c}$ the superconducting critical temperature
%which estimated as $5.66\times 10^{-4}$ eV.
%The superconducting gap is often used to compared with the excitation gap in normal region,
%thus we simply use $\Delta_{s}$ to replace the $\Delta_{s}+m_{D}^{\eta s_{z}\tau_{z}}$ in below.
%Obviously, the quasienergy spectrum here is distinct from the one obtained by the Floquet technique in low-momentum limit as presented in the
%Refs.\cite{López A,Wu C H5}.
%Note that since the exchange field considered here ($M_{s}=M_{c}=0.0039$ eV),
%the critical electric field (for the zero Dirac-mass) is no more 0.017 eV/\AA,
%but $\lesssim 0.051$ eV/\AA for small light-intensity. 

\section{Appendix: Computation of the Andreev bound level}

We focus on the computation of the dispersion of Andreev bound level in this section.
The Andreev level reads\cite{Wu C H7}
\begin{equation} 
\begin{aligned}
\varepsilon_{A}=s\frac{\Delta_{s}}{\sqrt{2}}\sqrt{1-\frac{A(C-{\rm cos}\phi)+s_{z}\sqrt{B^{2}[A^{2}+B^{2}-(C-{\rm cos}\phi)^{2}]}}{A^{2}+B^{2}}},
\end{aligned}
\end{equation}
where we have use the definitions
\begin{equation} 
\begin{aligned}
A=&C_{1}C_{2}+\frac{(S_{1}S_{2}(\frac{f_{2}}{f_{1}}+1)(\frac{f_{4}}{f_{3}}-1))}{4\sqrt{\hbar^{2}v_{F}^{2}k_{y}^{2}f_{2}/f_{4}+1}\sqrt{\frac{-f_{4}}{f_{3}}}\sqrt{-\hbar^{2}v_{F}^{2}k_{y}^{2}f_{1}/f_{3}+1}\sqrt{\frac{f_{2}}{f_{1}}}},\\
B=&\frac{S_{1}C_{2}(\frac{f_{3}}{2f_{1}}+\frac{1}{2})}{\sqrt{-(\hbar^{2}v_{F}^{2}k_{y}^{2}f_{1})/f_{3}+1}\sqrt{f_{2}/f_{1}}}-
\frac{C_{1}S_{2}(\frac{f_{4}}{2f_{2}}-\frac{1}{2})}{\sqrt{(\hbar^{2}v_{F}^{2}k_{y}^{2}f_{2})/f_{4}+1}\sqrt{-f_{4}/f_{3}}},\\
C=&\frac{\hbar^{2}v_{F}^{2}k_{y}^{2}S_{1}S_{2}}{\sqrt{\hbar^{2}v_{F}^{2}k_{y}^{2}f_{2}/f_{4}+1}\sqrt{-f_{4}/f_{3}}\sqrt{-(\hbar^{2}v_{F}^{2}k_{y}^{2}f_{1})/f_{3}+1}\sqrt{f_{2}/f_{1}}}\\
&-[1\cdot\Theta(\varepsilon_{n}-\mu_{n}-m_{D}^{\eta s_{z}\tau_{z}})+(-1)\cdot\Theta(-\varepsilon_{n}+\mu_{n}+m_{D}^{\eta s_{z}\tau_{z}}]\\
&\times\frac{(S_{1}S_{2}(f_{2}/f_{1}-1)(f_{4}/f_{3}+1))}{4\sqrt{\hbar^{2}v_{F}^{2}k_{y}^{2}f_{2}/f_{4}+1}\sqrt{-f_{4}/f_{3}}\sqrt{-(\hbar^{2}v_{F}^{2}k_{y}^{2}f_{1})/f_{3}+1}\sqrt{f_{2}/f_{1}}},
\end{aligned}
\end{equation}
with the Heaviside step function $\Theta$
which distinguish the two kinds of AR: retroreflection and specular AR,
and thus makes this expression valid for both of these two case.
The wave vectors $f_{1}\sim f_{4}$ and parameters $C_{1},\ C_{2},\ S_{1},\ S_{2}$ are defined as
\begin{equation} 
\begin{aligned}
f_{1}=m_{D}^{\eta\sigma_{z}\tau_{z}}+\varepsilon+\mu_{s},\ f_{2}=m_{D}^{\eta\sigma_{z}\tau_{z}}+\varepsilon-\mu_{s},\\
f_{3}=\varepsilon-m_{D}^{\eta\sigma_{z}\tau_{z}}+\mu_{s},\ f_{4}=m_{D}^{\eta\sigma_{z}\tau_{z}}-\varepsilon+\mu_{s},\\
C_{1}={\rm cos}(L\sqrt{f_{1}f_{3}/\hbar^{2}v_{F}^{2}-k_{y}^{2}}),\\
C_{2}={\rm cos}(L\sqrt{-f_{2}f_{4}/\hbar^{2}v_{F}^{2}-k_{y}^{2}}),\\
S_{1}={\rm sin}(L\sqrt{f_{1}f_{3}/\hbar^{2}v_{F}^{2}-k_{y}^{2}}),\\
S_{2}={\rm sin}(L\sqrt{-f_{2}f_{4}/\hbar^{2}v_{F}^{2}-k_{y}^{2}}),
\end{aligned}
\end{equation}
where $\mu_{s}$ is the chemical potential of the highly doped superconducting regime.
The $x$-component of the wave vector for the electron channel and hole channel, $k_{xe}$ and $k_{xh}$,
are incorporated in the above wave vectors,
specially, the electron and hole wave vectors here are both complex,
which implies the inclusion of the subgap solutions with the evanescent scattering waves,
%\cite{Rainis D},
%{http://gsfp.physi.uni-heidelberg.de/graddays_oktober_2009/content/en/zubehoer/anhaenge/dolcini/Supercond-Meso-Lecture-4.pdf}
and it has $\frac{\Delta_{s}}{2\varepsilon_{n}}2{\rm cos}\beta=1$ for $|\varepsilon_{n}|<\Delta_{s}$.
%but we only takes the real part during the computation as discussed below.
The $k_{y}\sim 2$ meV is conserved during our computation,
while the $k_{x}$ is unconserved during the scattering,
\begin{equation} 
\begin{aligned}
k_{x}^{e}=&\sqrt{\frac{f_{1}f_{2}}{\hbar^{2}v_{F}^{2}}-k_{y}^{2}},\\
k_{x}^{h}=&[1\cdot\Theta(\varepsilon_{n}-\mu_{n}-m_{D}^{\eta s_{z}\tau_{z}})+(-1)\cdot\Theta(-\varepsilon_{n}+\mu_{n}+m_{D}^{\eta s_{z}\tau_{z}}]
\sqrt{\frac{f_{4}(-f_{2})}{\hbar^{2}v_{F}^{2}}-k_{y}^{2}}.
\end{aligned}
\end{equation}
That's similar to the result of Ref.\cite{Beiranvand R} which is for graphene and thus with valley degenerates:
\begin{equation} 
\begin{aligned}
k_{x}^{e(h)}=\frac{1}{\hbar v_{x}}(\mu+(-)\varepsilon)\sqrt{1+\frac{2\eta\lambda_{SOC}}{\mu+(-)\varepsilon}}
\sqrt{\frac{\hbar v_{y}^{2}k_{y}^{2}}{\hbar v_{x}^{2}k_{x}^{e(h)2}}+1}
\end{aligned}
\end{equation}
in the presence of Rashba-coupling.
%Note that we only consider the dc Josephson effcet in the thermodynamic equilibrium state
%here rather than the ac Josephson effcet which with time-dependent phase difference
%(e.g., by a time-dependent bias voltage).

\end{large}
\renewcommand\refname{References}

\clearpage
%\section{Table}
%Table 1:Three kinds of distorted structure of silicene which are transferred into the insulating phase with unambiguous band gaps
%compare to the pristine one.
%The corresponding bond angles between e-bond and f-bond and the (average) bulkling distances $\overline{\Delta}$ are also shown.
%The corresponding schematic of this displacing (distorting) was shown in the Fig.4(e).
%\begin{table}[!hbp]
%\centering
%%\resizebox{\textwidth}{!}{
%%\begin{threeparttable}
%%\begin{spacing}{1.19}
%\begin{tabular}{cccccccccc}
%\hline
%\hline  
%structure                   &  a (\AA)   &b (\AA)&c (\AA)&d (\AA)&e (\AA)&f (\AA)&$\overline{\Delta}$ (\AA) &Bond angle&Band gap (eV)\\
%\hline    
%pristine &2.280 &2.280 &2.280 &2.280 &2.280  &2.280 &0.51&115.96$^\text{o}$& $5.45\times 10^{-7}$\\
%distorted 1    &  2.245    &2.245&2.263&2.263&2.246&2.247&0.47 &116.821$^\text{o}$&1.609\\
%distorted 2    &  2.245    &2.245&2.263&2.263&2.245&2.246 &0.45&117.119$^\text{o}$&1.610\\
%distorted 3    &  2.245    &2.245&2.264&2.263&2.245&2.245 &0.42&117.247$^\text{o}$&1.612\\
%\hline
%\hline     
%\end{tabular}
%%\end{spacing}
%%\end{threeparttable}}
%\end{table}

\clearpage
Fig.1
\begin{figure}[!ht]
   \centering
 \centering
   \begin{center}
     \includegraphics*[width=0.9\linewidth]{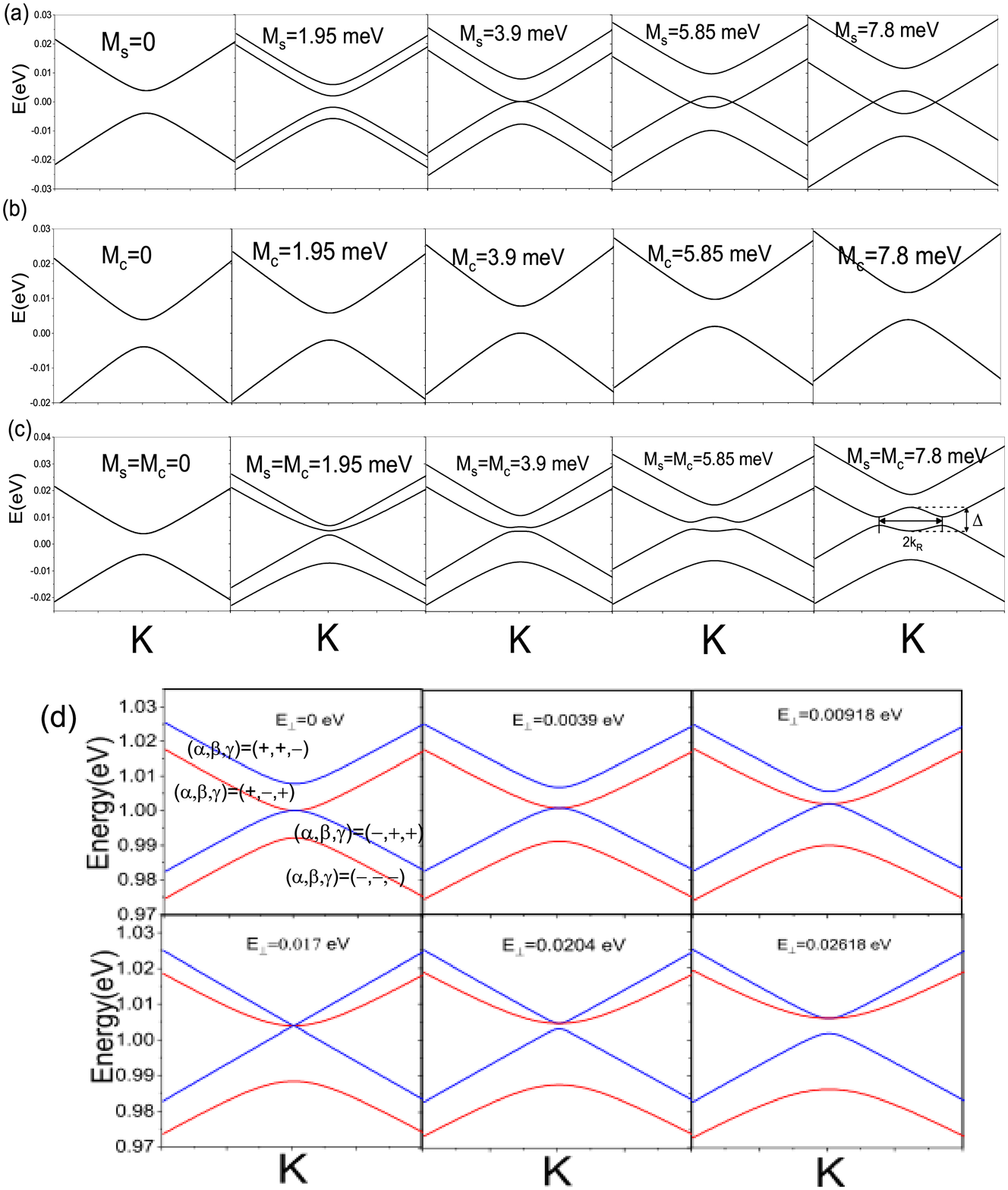}
\caption{Band gap evolution in valley K under the strength of of $M_{s}$ (a), $M_{c}$ (b), and both of them (c).
Both the external fields and the on-site intraction $U$
are setted as zero here.
(d) is the band gap evolution in valley K with the intrinsic SOC and NNN Rashba-coupling $\lambda_{R_{2}}$\cite{Wu C H3}.
The avoided corssing effect is obvious in the last panel of (c).
In (d), the index $(\alpha,\beta)$ are explanined in the text,
the index $\gamma$ denotes the spin helical;
$\gamma=-$ for anticlockwise spin helical and $\gamma=+$ for clockwise spin helical.
}
   \end{center}
\end{figure}
\clearpage

Fig.2
\begin{figure}[!ht]
   \centering
 \centering
   \begin{center}
     \includegraphics*[width=0.9\linewidth]{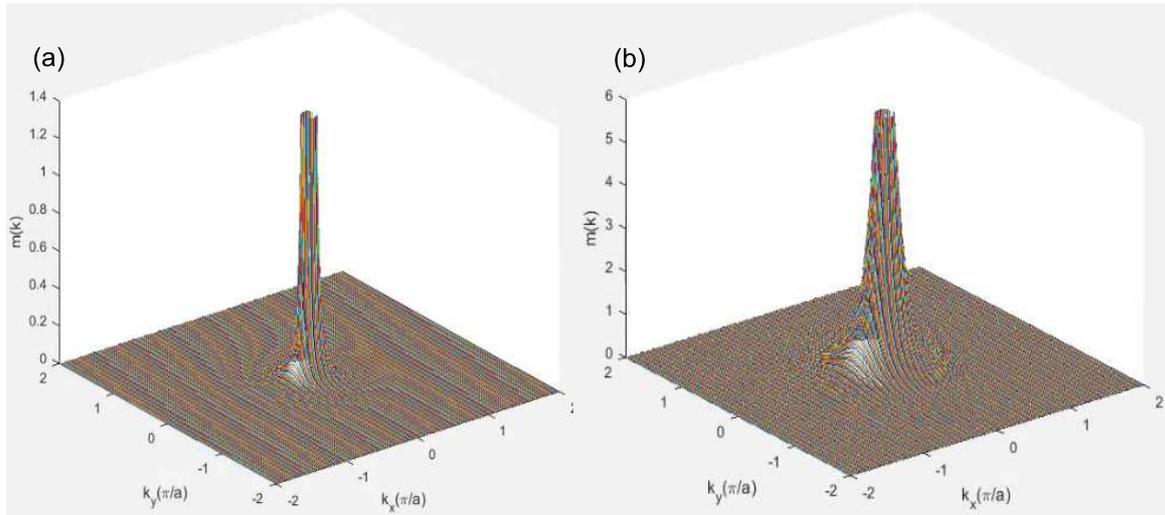}
\caption{(Color online) Orbital magnetic moment $m({\bf k})$ in the two dimension momentum space for Dirac mass 
$m_{D}^{\eta \sigma_{z}\tau_{z}}=0.005$ eV (a) and $m_{D}^{\eta \sigma_{z}\tau_{z}}=0.32$ eV (b).
The vertical axis is in unit of $e/\hbar$.
}
   \end{center}
\end{figure}

%\clearpage
%Fig.1
%\begin{figure}[!ht]
%\subfigure{
%\begin{minipage}[t]{0.5\textwidth}
%\centering
%\includegraphics[width=0.9\linewidth]{orbitalmagneticmoment.eps}
%\label{fig:side:a}
%\end{minipage}
%}
%\subfigure{
%\begin{minipage}[t]{0.5\textwidth}
%\centering
%\includegraphics[width=0.9\linewidth]{berrycurvature.eps}
%\label{fig:side:b}
%\end{minipage}
%}\\
%\subfigure{
%\begin{minipage}[t]{0.5\textwidth}
%\centering
%\includegraphics[width=0.9\linewidth]{berry2.eps}
%\label{fig:side:a}
%\end{minipage}
%}
%\subfigure{
%\begin{minipage}[t]{0.5\textwidth}
%\centering
%\includegraphics[width=0.9\linewidth]{berry3.eps}
%\label{fig:side:b}
%\end{minipage}
%}
%\caption{(Color online)
%Plots of the orbital magnetic moment (left-top panel) and the Berry curvature (right-top panel and bottom panels) for the conduction band in valley K'
%(the another valley K which is antisymmetry with K is not shown).
%The datas adoped are as following: the peak of $m(k)$ is center in the $k_{x}=2.092\ \pi/a$, $t=1.6$ eV, band gap $\Delta=0.01$ eV and $\Delta=0.2192$ eV
%and $\Delta=-0.2192$ eV,
%chemical potential $\mu=0.1$ eV. The unit of $\hbar=c=1$ are used.}
%{Valley-Contrasting Physics in Graphene Magnetic Moment and Topological Transport}
%\end{figure}

\clearpage
Fig.3
\begin{figure}[!ht]
   \centering
 \centering
   \begin{center}
     \includegraphics*[width=0.3\linewidth]{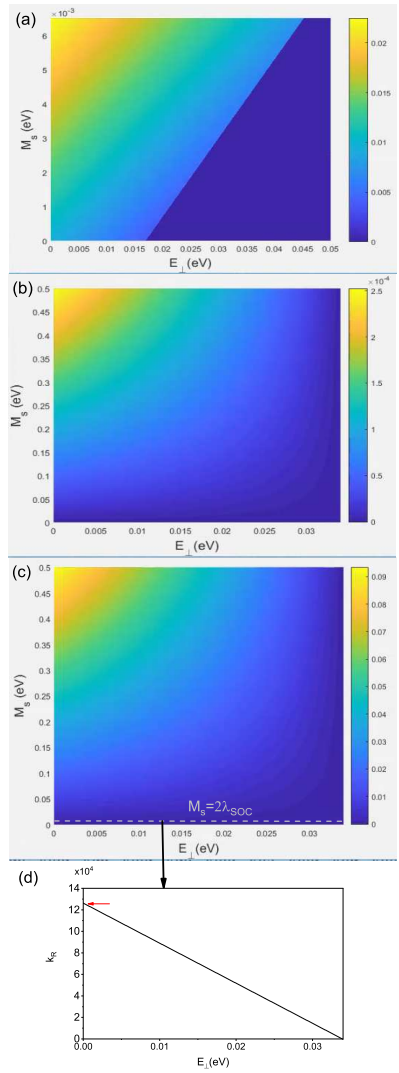}
\caption{(Color online) (a) is the band gap $\Delta$ (in unit of eV) in the case of $|\lambda_{SOC}-\frac{\overline{\Delta}}{2}E_{\perp}+M_{s}|\ge 0$.
(b) is the band gap (in unit of eV) created by the anticrossing of bands, such band gap enlarge with the increase of $M_{s}$,
and decrease with the increase of $E_{\perp}$,
it's corresponding radiu $k_{R}$ is presented in the (c).
Here we set $\sigma_{z}=-1$, $\tau_{z}=-1$ for (a) and $\sigma_{z}=1$, $\tau_{z}=-1$ for (b) and (c).
(d) shows the enlarged view at $M_{s}=2\lambda_{SOC}=0.0078$ eV (i.e., the gray dash-line in (c)),
which is precisely the case of the last panel of Fig.1(c).
The red arrow indicates the radiu
of quantum anomalous Hall phase ($M_{s}=2\lambda_{SOC}$, $E_{\perp}=0$) is about ${\bf k}_{R}=0.00126$ (i.e., the one labeled in the Fig.2(c)),
and the radiu ${\bf k}_{R}$ for $M_{s}=2\lambda_{SOC}$ vanish at $E_{\perp}=\frac{4\lambda_{SOC}}{\overline{\Delta}}\approx 0.0339$ eV.
}
   \end{center}
\end{figure}
\clearpage
Fig.4
\begin{figure}[!ht]
   \centering
 \centering
   \begin{center}
     \includegraphics*[width=0.3\linewidth]{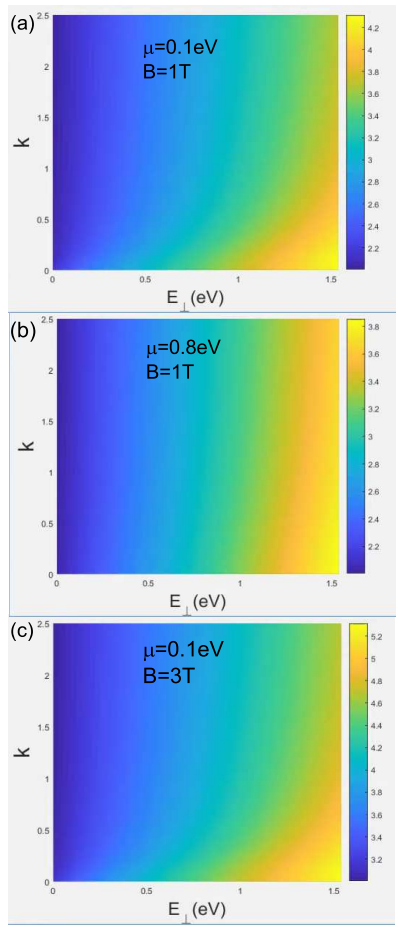}
\caption{Momentum operator $\mathcal{P}$ about the interband processin $E_{\perp}-{\bf k}$ space
with different chemical potential and magnetic field.
Note that here we set $c=m_{0}=\hbar=v_{F}=1$ for simplicity,
and to obtain the result efficiently,
the Berry curvature is also treat as 1 to avoid the large peak the the small ${\bf k}$ regime (as shown in the Fig.2).
}
   \end{center}
\end{figure}
\clearpage

Fig.5
\begin{figure}[!ht]
   \centering
 \centering
   \begin{center}
     \includegraphics*[width=0.8\linewidth]{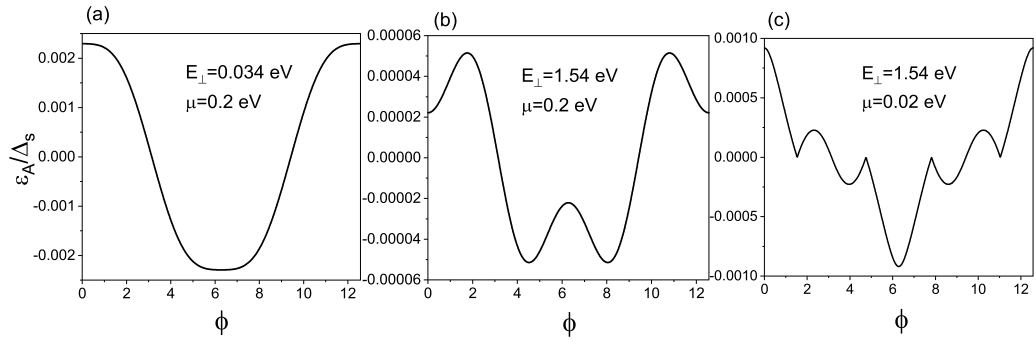}
\caption{Andreev bound state level versus the phase difference for $E_{\perp}=0.034$ eV, $\mu=0.2$ eV (a) and 
$E_{\perp}=1.54$ eV, $\mu=0.2$ eV (b) and
$E_{\perp}=1.54$ eV, $\mu=0.02$ eV (c).
The range of $\phi$-axis is setted as one period, $4\pi$.
The intensity of off-resonance light is setted as 0.3\cite{Wu C H5,Wu C H7}.
}
   \end{center}
\end{figure}
\clearpage

\end{document}